\documentclass[journal]{IEEEtran}
\usepackage[utf8]{inputenc}

\makeatletter
\def\endthebibliography{%
  \def\@noitemerr{\@latex@warning{Empty `thebibliography' environment}}%
  \endlist
}
\makeatother

\IEEEoverridecommandlockouts
\usepackage{cite}
\usepackage{amsmath,amssymb,amsfonts}
\usepackage{algorithmic}
\usepackage{graphicx}
\usepackage{textcomp}
\usepackage{xcolor}
\usepackage{color}
\usepackage{comment}
\usepackage{multirow}
\usepackage{ bbold }
\usepackage{booktabs}
\usepackage{hhline}
\usepackage{tabularx}

\usepackage{tikz}
\usetikzlibrary{arrows}
\newcommand*\circled[1]{\tikz[baseline=(char.base)]{
            \node[shape=circle,draw,inner sep=1pt] (char) {#1};}}

\definecolor{cadmiumgreen}{rgb}{0.0, 0.42, 0.24}

\newcommand{\AH}[1]{\textcolor{cyan}{{#1}}}

\usepackage{makecell}
\def\BibTeX{{\rm B\kern-.05em{\sc i\kern-.025em b}\kern-.08em
    T\kern-.1667em\lower.7ex\hbox{E}\kern-.125emX}}

\begin{document}


\title{Human readable network troubleshooting\\based on anomaly detection and feature scoring}

\author{
\IEEEauthorblockN{Jose M. Navarro,  Alexis Huet, Dario Rossi}\\
\IEEEauthorblockA{Huawei Technologies Co. Ltd.}\\
\texttt{\{jose.manuel.navarro, alexis.huet, dario.rossi\}@huawei.com}
}

\maketitle
\thispagestyle{empty}
\begin{abstract}
Network troubleshooting is still a heavily human-intensive process. To reduce the time spent by human operators in the diagnosis process, we present a system based on (i) \emph{unsupervised learning} methods for detecting anomalies in the time domain, (ii)  an \emph{attention mechanism} to rank features in the feature space and finally (iii) an \emph{expert knowledge} module able to seamlessly incorporate previously collected domain-knowledge.

In this paper, we thoroughly evaluate the performance of the full system and of its individual building blocks: particularly, we consider (i) 10 anomaly detection algorithms as well as (ii) 10 attention mechanisms, that comprehensively represent the current state of the art in the respective fields. Leveraging a unique collection of expert-labeled datasets worth several months of real router telemetry data, we perform a thorough performance evaluation   contrasting practical results in constrained stream-mode settings, with the results achievable by an ideal oracle in academic settings.

 Our experimental evaluation shows that (i) the proposed system is effective in achieving high levels of agreement with the expert,
 and  (ii) that even a simple statistical approach is able to extract  useful information from expert knowledge gained in past cases, significantly  improving troubleshooting performance.
\end{abstract}

\begin{IEEEkeywords}
Anomaly detection;  Stream learning;   Unsupervised learning;   Network monitoring;    Model driven telemetry; 
\end{IEEEkeywords}

\section{Introduction}
Network operation and management, especially concerning troubleshooting, is still a largely manual and time-consuming process\cite{mazel2011sub}. While keeping human operators ``in the loop'' is unavoidable nowadays, e.g. for accountability,
the use of machine learning techniques  can help removing humans from the ``fast loop'', by  automatically processing large volumes of data to replace human tasks~\cite{dromard2016online,bhatia2019unsupervised},  and overall increasing the troubleshooting efficiency. 
Additionally, keeping human operators in the ``slow loop'' can be beneficial, e.g., by providing rare labels to improve the 
machine learning algorithms.
Under these premises, it becomes obvious that the amount of time spent by human experts in the troubleshooting process is a  valuable resource that anomaly detection systems should explicitly take into account~\cite{sota_human_labelless,sota_human_opprentice,sota_human_sfe,sota_human_feedback}, which is our focus in this work. As such, it is clear that the ultimate goal of an anomaly detection system is to produce a ``human readable'' output~\cite{sota_other_adele,sota_ad_dl1, sota_cloudDet,sota_refout,sota_beam,sota_lookout,sota_hics,PROTEUS}
 that should, furthermore, be presented to the final users in a way that minimally alters their workflow.

\begin{figure*}[t]
\begin{center}
  \includegraphics[width=1\textwidth]{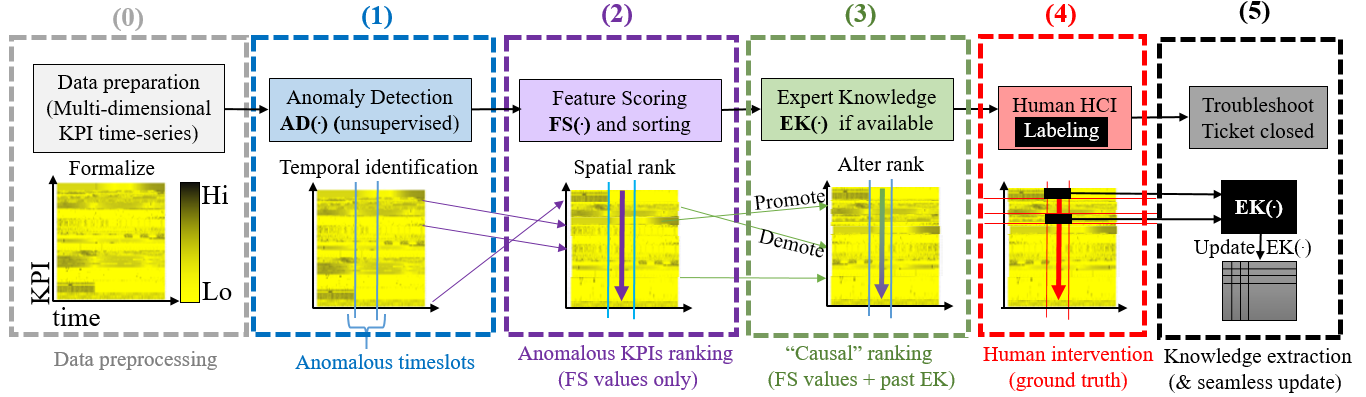}
  \caption{High-level view of the building blocks of employed  in the Human Readable Router Anomaly Detection (HURRA) system presented in this paper.}
  \label{fig:birdeye}
\end{center}
\end{figure*}

We design a system for Human Readable Router Anomaly  (HURRA) detection by combining simple building blocks, that are depicted from a very high-level perspective in  Fig.~\ref{fig:birdeye}.
From the left, router KPI data, which can be abstracted as multi-variate time series, is the input to a  \circled{0} preliminary \textit{data preprocessing} step: for instance, the heat map representation of one of our datasets shown in Fig.~\ref{fig:birdeye}, employs a standard feature regularization.
Next, \circled{1} an \textit{anomaly detection (AD)} algorithm is used to identify, along the temporal dimension, time-regions where the system  behavior presents unexpected patterns. 
Subsequently, an \circled{2} attention-focus mechanism, implemented as a  \textit{feature scoring (FS)} policy is used to rank KPIs along the spatial dimension. The purpose is to only minimally alter the existing troubleshooting system, by letting users prioritize their focus to KPIs that are likely the most relevant for troubleshooting the issue at hand.  
 
To further assist  human intervention, HURRA additionally leverages, if available,  \circled{3}  previous \textit{expert knowledge (EK)}  with a simple mechanism based on statistical information that alters the KPI ranking, using separate schemes to promote/demote KPIs in the ranking. By prioritizing KPIs that  experts have found to be relevant in the solution of previous troubleshooting tickets, the EK mechanism attempts at presenting KPIs that network operators more often associate with what they loosely identify as the ``root cause''  of the issue (we point out that we are not explicitly making use of causality theory\cite{pearl}, hence the quotes).  By \circled{4} explicitly flagging KPIs, experts (unconsciously) provide \textit{ground truth labels} to our system: from these labels, it is possible to \circled{5} automatically distill and \textit{update expert knowledge}, without requiring any further human interaction.

We benchmark our proposal against 64 datasets comprising  manually labeled troubleshooting data, gathered from several operational ISPs deployments, with minute-level telemetry.  Intuitively, the main objective of HURRA is to obtain high agreement between the ranking of scores produced by the expert vs the algorithmic output, which is well captured by metrics such as the normalized Discounted Cumulative Gain (nDCG)~\cite{yahoo_rankings_from_xgboost}.
Summarizing our main contributions:
\begin{itemize}
\item We design a fully unsupervised system achieving a high agreement with expert labeling  (nDCG up to 0.90 in our dataset), reducing the number of KPIs to be manually verified (16 KPIs or 4-fold improvement over an ideal anomaly detection Oracle missing our proposed FS method).

\item We propose a simple yet effective mechanism to account for expert knowledge~\cite{patent} that is incremental by design and that can further improve agreement with the expert -- even when an ideal Oracle is used to tackle the anomaly detection problem.

\item We conduct an extensive performance evaluation including 10 AD algorithms and 10 FS strategies -- covering both practical results from deployment conditions, as well as their distance from the optimal performance gathered in ideal academic settings.  
\end{itemize}

The rest of this paper overviews related work (Sec.~\ref{sec:related}),
describes the dataset (Sec.~\ref{sec:dataset}), details our methodology (Sec.~\ref{sec:methodology}). Experimental result of the AD, FS  and EK blocks (Sec.~\ref{sec:results:AD}--Sec.~\ref{sec:results:FS}) are reported next. We finally discuss performance of the full blown system  (Sec.~\ref{sec:results:E2E}) and summarize our findings   (Sec.~\ref{sec:conclusions}).

\section{Related work}\label{sec:related}
HURRA does not attempt to make any novel contribution for building blocks \circled{0}-\circled{1} of Fig.~\ref{fig:birdeye}, for which we nevertheless carry on a systematic exploration of  state-of-the art techniques,  an original contribution which is interesting per se. 
HURRA proposes  an \circled{2} attention focus mechanism that can seamlessly integrate \circled{3} expert knowledge, and leverage a unique \circled{4} human labeled datasets for evaluation purposes. In this section, we separately overview the state of the art pertaining to anomaly detection (Sec.~\ref{sec:related:ad}) and feature scoring (Sec.~\ref{sec:related:fs}).

\subsection{Anomaly detection}\label{sec:related:ad}
This section presents a taxonomy of the main anomaly detection methods that can be used as the building block \circled{1} of HURRA.  As anomalies are rare event, there is scarcity of labels, for which \emph{unsupervised algorithms} are preferable.
At high-level, AD methods take as input a multivariate series and outputs a score for each timestamp such that the larger the score the more anomalous is the timestamp. 

A binarization step is then applied to convert the scores vector into a binary vector, $1$ corresponding to an anomalous timestamp and $0$ to a normal one.
For some algorithms, the binarized result is directly provided without scores.

Tab.~\ref{table:algorithms} summarizes the main unsupervised anomaly detection algorithms and their main characteristics. Particularly, while in the preliminary version of this work~\cite{itc32} we limitedly considered two \emph{batch} methods (namely DBSCAN~\cite{dbscan} and IF~\cite{sota_ad_isofor}), in this extended version we systematically evaluate a significantly larger set of algorithms, that comprehensively represent the state of the art in the Knowledge and Data Discovery (KDD) field, and that we briefly overview in this section. 
Given that no single unsupervised algorithm is known to perform well in any settings, our aim is to guide the selection of such an algorithm for the network domain in a comprehensive and rigorous way.

\newcommand{\doublecheck}[0]{\checkmark\kern-0.3em\checkmark}

\begin{table}[!t]
\caption{Taxonomy of main anomaly detection algorithms. }
\label{table:algorithms}
\centering
\begin{tabularx}{0.5\textwidth}{lllll}
 {\bf Class} &{\bf (Year) Algorithm}        &  {\bf Method}  &  {\bf Processing} & \\ 
\hline
\parbox[t]{2mm}{\multirow{10}{*}{\rotatebox[origin=c]{90}{\textbf{Proximity-based}}}}
 &(1951) kNN~\cite{knn1951,ramaswamy2000efficient_KNNth_AD}              & Distance        &     Batch        &                                 \\ 
 &(1998) DB-outlier~\cite{knox1998algorithms_distancebasedoutlier} & Distance &  Batch & \\ 
 &(2007) STORM~\cite{angiulli2007detecting_STORM}       & Distance       &  Stream & \\  
\cline{2-5}
 &(1996) DBSCAN~\cite{dbscan}                  &  Density &     Batch     & \doublecheck          \\ 
 &(2000) LOF~\cite{breunig2000lof_theoriginal}   & Density        &    Batch      & \checkmark          \\ 
 &(2003) LOCI~\cite{papadimitriou2003loci}  & Density        & Batch &                                 \\
\cline{2-5}
 &(2003) CBLOF~\cite{he2003discovering_CBLOF} & Clustering & Batch  & \\ 
 &(2011) MCOD~\cite{kontaki2011continuous_MCOD}             & Clustering        &  Stream          &   \checkmark                               \\ 
 &(2020) OADDS~\cite{putina2020online_OADDS}     & Clustering       &   Stream         & \checkmark          \\ \hline

\parbox[t]{2mm}{\multirow{8}{*}{\rotatebox[origin=c]{90}{\textbf{Ensemble-based}}}}
 &(2008) IF~\cite{sota_ad_isofor}           & Trees        & Batch        & \doublecheck         \\
 &(2011) HST~\cite{tan2011fast_HST}         & Trees       &    Stream     & \checkmark          \\ 
 &(2016) RRCF~\cite{guha2016robust_RRCF}    & Trees       &  Stream  & \checkmark          \\ 
 &(2020) RHF~\cite{ICDM-20_RHF}             & Trees        &  Batch & \checkmark          \\ 
\cline{2-5}
 &(2005) Bagging~\cite{lazarevic2005feature_bagging} & Subspace & Batch & \\ 
 &(2016) RS-Hash~\cite{sathe2016subspace_RS_hash} & Subspace & Stream/Batch & \\
 &(2016) LODA~\cite{pevny2016loda}    & Subspace &  Stream/Batch   & \checkmark  \\
 &(2018) xStream~\cite{manzoor2018xstream} & Subspace & Stream    & \checkmark   \\ 
 \hline
\multicolumn{5}{l}{}\\
\multicolumn{5}{l}{\checkmark \emph{the algorithm is used in the experimental section.}}\\
\multicolumn{5}{l}{\doublecheck \emph{the algorithm was also used in the preliminary conference version\cite{itc32}.}}
\end{tabularx}
\end{table}

Depending on the algorithm, the processing is done either \emph{in batch}, where the full dataset is took as a static input, or \emph{in streaming}, where the dataset is treated sequentially over time. While batch algorithms only focus on identifying outliers in the global distribution of the KPIs, streaming algorithms have the advantage of explicitly taking into account the temporal dimension. For instance they are able to detect local anomalies, which are values appearing anomalous compared to  temporally close values, even though they are not necessarily outliers in the global distribution.   At high level, algorithms can be divided in two main classes of  \emph{proximity-based} vs \emph{ensemble-based} algorithms, that we further outline next. \\

\subsubsection{Proximity}
In turn, proximity-based algorithms can be  categorized 
depending on the method used to detect anomalies, by either measuring the \emph{distance} between datapoints, evaluating the \emph{density} surrounding datapoints, or aggregating together close datapoints into \emph{clusters}.

\paragraph{Distance}
Distance-based algorithms are the oldest, with classic algorithms such as K-Nearest Neighbors (kNN)  (kNN)~\cite{knn1951}, where each datapoint is scored with the distance to its $k$-nearest neighbors, of which numerous variants~\cite{ramaswamy2000efficient_KNNth_AD,angiulli2002fast_KNN_AD} were developed over time. Similarly, the Distance-Based Outlier (DB-Outlier)~\cite{knox1998algorithms_distancebasedoutlier}  algorithm flags a sample as anomalous if the distance needed to reach a proportion $p$ of the other datapoints is greater that $R$. 
STream OutlieR Miner (STORM)~\cite{angiulli2007detecting_STORM}  extends those previous distance-based algorithms to the streaming case, by introducing a sliding window over time. Exact-STORM considers that a datapoint is anomalous if, among the $W$ last points corresponding to the window, the number of neighbors at distance $R$ or less is smaller than $k$, and approximated variants are proposed to reduce memory requirement~\cite{angiulli2007detecting_STORM}.
In high dimensional datasets, distance computation becomes however a crucial  weaknesses, both 
due to the ``curse of dimensionality''~\cite{beyer1999nearest_curse_dimensionality} and  in terms of computational complexity, for which we avoid distance-based methods.

\paragraph{Density} To circumvent the above problems, researchers started then  using low-density regions for identifying outliers. The most well known algorithm in this category is Density-based Spatial Clustering of Applications with Noise  (DBSCAN)~\cite{dbscan}, which first identifies core points based on proximity (as for KNN, a core point has at least $k$ neighbors at distance not bigger than $R$), before deducing outlier points as those which cannot reach any core point (i.e., in low density region from all core points).
Local Outlier Factor (LOF)~\cite{sota_ad_lof} extends DBSCAN by introducing local densities: for each datapoint $x$, a local reachability density number is computed by measuring how far to $A$ are the $k$-nearest neighbors of $x$ (measured with the asymmetric concept of reachability). Then the local reachability density of $x$ is compared to that of its neighbors, obtaining a  score  measuring the degree of anomaly (a value above one corresponds to a datapoint in a region with lower density compared to neighbors).
Taking a complementary approach, Local Correlation Integral (LOCI)~\cite{papadimitriou2003loci}
 compares, for all possible values of $R$, a deviation in the number of neighbors found in region $R$ and $\alpha R$ around each datapoint (where $\alpha < 1$ is a free hyperparameter), at the price of a higher complexity. It follows that DBSCAN and LOF are good choices for batch-mode proximity-based algorithms.

\paragraph{Clustering}
More recent, proximity-based algorithms explicitly consider a clustering step to identifying the anomalous samples. For instance, Cluster-based Local Outlier Factor (CBLOF)~\cite{he2003discovering_CBLOF} first apply any clustering algorithm with $k$ clusters (e.g. $k$-means), decides whether each cluster is large or small based on two parameters $\alpha$ and $\beta$, before finally scoring each datapoint based both on the size of the corresponding cluster and on the distance from the datapoint to a large cluster centroid. 
More advanced algorithms~\cite{cao2006density_denstream,kontaki2011continuous_MCOD,putina2020online_OADDS} push further the notion of micro-clusters. 
Both Micro-cluster Continuous Outlier Detection (MCOD)~\cite{kontaki2011continuous_MCOD} 
and  Outlier Anomaly Detection  for Data Streams (OADDS)~\cite{putina2020online_OADDS}, update an online sliding window over time,  aggregating and updating  samples into micro-clusters, and
flagging points that fall outside any micro-cluster as anomalies. 
They however differ in the way the updates are performed and the clusters are managed: instead of a fixed window as in STORM or MCOD, OADDS leverages the DenStream~\cite{cao2006density_denstream} clustering algorithm, using an exponentially fading  window and aggregates datapoints into microclusters not only in the space dimension but also in the time one.  As such, MCOD and OADDS  are both worth considering as representative of stream-mode proximity-based algorithms.\\

\subsubsection{Ensembles}

Ensemble-based algorithms can be further divided into subcategories: \emph{tree}-based ensembles isolate datapoints at the leaf of a tree and combine many such weak learners to give a robust result, whereas \emph{subspaces}-based methods repeatedly extracts subspaces (or projects the datapoints into low-dimensional subspaces) to identify datapoints of sparsely populated regions. 

\paragraph{Trees} Tree-based ensembles differ (i) in the way trees are built, and (ii) in the  extraction of the anomalous scores from the forest.  Isolation Forest (IF)~\cite{sota_ad_isofor},  by far most popular algorithm in this category, build each tree by taking an initial root node containing all datapoints, and then by recursively partitioning the datapoints related to each node into two child nodes $\lbrace \text{KPI} \leq x \rbrace$ and $\lbrace \text{KPI} > x \rbrace$, with $\text{KPI}$ and $x$ randomly (uniformly) selected. 

On this tree, each datapoint is associated with a leaf node, and the path length from the root is measured: the underlying idea is that we expect that a shorter path corresponds to a more anomalous datapoint,  being easily isolated from the others. Robust results is obtained by computing the  average path length over multiple trees.
Random Histogram Forest (RHF)~\cite{ICDM-20_RHF} is another batch algorithm that builds trees recursively  with a split that is not uniform but weighted on the the fourth moment of the feature (i.e., the kurtosis, used to to favour the selection of features containing outliers): the depth of the tree is set by a parameter, and anomalous score is then computed as a function of the number of points in the leaves.
Half-Space Trees (HST)~\cite{tan2011fast_HST} builds trees in streaming mode (benefiting from a simple procedure for updating the scores), with a cut defined as the middle of the range of the selected feature (instead of randomly picked as in IF).
Robust Random Cut Forest (RRCF)~\cite{guha2016robust_RRCF} is another streaming algorithm based on IF, where the anomalous score  depends on the collusive displacement, i.e., how much the complexity of the tree has changed before and after adding the datapoint.
Tree-based ensembles such as IF, RHF, HST and RRCF are currently considered as the state of the art for anomaly detection, and as such are worth including systematically.

\paragraph{Subspace}
Subspace-based methods represent the most recent category of anomaly detection algorithms. The main ides behind subspaces is to combat the curse of dimensionality  and  deal with irrelevant or noisy KPIs by applying an algorithm on multiple subspaces and combining the outputs into a single ensemble score.  Feature bagging~\cite{lazarevic2005feature_bagging}  is a popular and generic subspace strategy, which consists of applying an anomaly detection algorithm (e.g. LOF) on different random subspaces before taking the average of the scores over the repetitions. 
Among the most relevant and recent approaches, 
Randomized Subspace Hashing algorithm (RS-Hash)~\cite{sathe2016subspace_RS_hash} mostly follows the feature bagging template, but differs in the inner algorithm and particularly in the specific hash function  provided to manage the streaming case.
Lightweight on-line detector of anomalies (LODA)~\cite{pevny2016loda} is one of the few streaming algorithms based on random projections to low dimensional random spaces, aggregating histograms over subspaces to compute anomalies scores in the original domain. 
xStream\cite{manzoor2018xstream}  introduces new data structures (half-space chains) and
outlier score calculation (by improving the histogram projections by, e.g., including an efficient multi-scale counting and the addition of random shifts to avoid splitting data clusters at low-scales). Notably, it is 
capabale of handling streams of variable dimensionality and missing data, which makes it robust from a practical viewpoint.
LODA and xStream represent the most recent state of the art in stream-mode subspace-based methods and are thus promising candidates worth including in the comparison.

\subsection{Feature scoring}\label{sec:related:fs}

Admittedly, as summarized in Tab.\ref{tabl:sota:fs} we are not the first attempting
at ranking features~\cite{sota_other_adele,sota_ad_dl1, sota_cloudDet}, or explaining outliers~\cite{sota_refout,sota_beam,sota_lookout,sota_hics,PROTEUS},   possibly leveraging  
expert knowledge~\cite{sota_human_labelless,sota_human_opprentice,sota_human_sfe,sota_human_feedback}.

\begin{table}[!t]
\caption{Taxonomy of feature scoring algorithms. }
\label{tabl:sota:fs}
\centering
\begin{tabularx}{0.5\textwidth}{p{0.3cm}lllll}
\toprule
 {\bf Class$^\dagger$} &{\bf (Year) Algorithm}         &  {\bf Method} &  {\bf Granularity} \\ 
\midrule
  &(2018) Normal Distribution~\cite{sota_other_adele}                  &  Parametric & Point-based & \checkmark                              \\ 
  &(2019) Reconstruction error~\cite{sota_ad_dl1}     &  Parametric & Point-based& \\ 
 \textbf{FR} &(2019) STL decomposition~\cite{sota_cloudDet}      &  Parametric & Point-based& \\
  &(2020) Average Deviation~\cite{itc32} &  Non-param & Group-based &  \doublecheck \\
  &(2020) Rank Deviation~\cite{itc32} &  Non-param  & Group-based  & \doublecheck \\
 \hline
  &(2013) RefOut ~\cite{sota_refout}     &  Non-param  & Point-based &   \\ 
  &(2016) Beam~\cite{sota_beam}     &  Non-param & Point-based &   \\ 
 \textbf{OE} &(2018) Lookout~\cite{sota_lookout} &  Non-param & Group-based & \\ 
  &(2012) HiCS~\cite{sota_hics}     &  Non-param & Group-based &  \\
  &(2021) PROTEUS~\cite{PROTEUS} & Parametric & Group-based  &\checkmark \\
\bottomrule
\multicolumn{4}{l}{}\\
\multicolumn{4}{l}{$^\dagger$\emph{\textbf{FR} = Feature Ranking. \textbf{OE} = Outlier Explanation.}}\\
\multicolumn{4}{l}{\checkmark \emph{used in the experimental section.}}\\
\multicolumn{4}{l}{\doublecheck \emph{also used in the preliminary conference version\cite{itc32}.}}
\end{tabularx}
\end{table}

\subsubsection{Feature ranking} 
Our work proposes two non-parametric, simple yet effective feature raking strategies, based on average or rank deviation, that we limitedly compared  to na\"ives FS strategies (Random, Alphabetical) in the preliminary version of this work\cite{itc32}. Therefore, we extend the comparison by including relevant model-based alternatives such as~\cite{sota_other_adele}, which ranks features based on how much they deviate from a normal distribution.
 
We point out that LSTM and autoencoders are instead exploited  in \cite{sota_ad_dl1}, where reconstruction errors
are interpreted as anomalies: while this scheme is interoperable with our,
the temporal extent and heterogeneity of the data at our disposal (cfr. Sec.~\ref{sec:dataset}) do not allow us to properly train a deep neural network.
CloudDet~\cite{sota_cloudDet} instead detects anomalies through scoring  seasonally decomposed components of the time series: this couples \circled{1} and \circled{2} by assigning and scoring anomalies to each feature independently, which limits the analysis to a set of univariate features, and is thus orthogonal to our goal.

\subsubsection{Outlier explanation} Outlier explanation attempt to reduce the feature space into a smaller dimension that best represent the outlier by either (i) random subspacing \cite{mazel2011sub,sota_refout,sota_beam,sota_lookout,sota_hics}
or (ii) exploiting predictive power of each dimension \cite{PROTEUS} 
and can either tackle single (\emph{point explanation}  \cite{sota_refout,sota_beam}) or multiple  outliers (\emph{summarization explanation}~\cite{sota_lookout,sota_hics,PROTEUS}).
The interest for subspace methods in network domain was first raised in  \cite{mazel2011sub}, which however mostly focuses on anomaly detection, instead of explanation\cite{sota_refout,sota_beam,sota_lookout,sota_hics}.  
For instance, RefOut~\cite{sota_refout} performs point explanation by creating random subspaces, assessing feature importance by comparing the distribution of the scores produced by an AD algorithm. Likewise, Beam~\cite{sota_beam} greedily builds the most interesting subspace through the maximization of the AD scores, while  LookOut~\cite{sota_lookout} and HiCS~\cite{sota_hics} extend this approach by a more  complex  characterization of  the interaction among features. 

Complementary approach such as PROTEUS~\cite{PROTEUS}, tackles the problem by creating a \emph{predictive explainer} via feature selection and supervised classification algorithms, in which the feature importance in the supervised model  is used to explain future anomalies.
Despite some limits (e.g.,  high computational cost, or the assumption of sample independence which does not hold for time series) these algorithms are interesting: we thus include four predictive methodologies based on \cite{PROTEUS} as an ideal comparison point to our proposals.

\newcommand{\superquad}[0]{$\qquad\qquad\qquad\qquad\qquad\qquad\qquad\qquad$}
 \begin{figure*}[!t]
 \centering
  \includegraphics[width=0.325\textwidth]{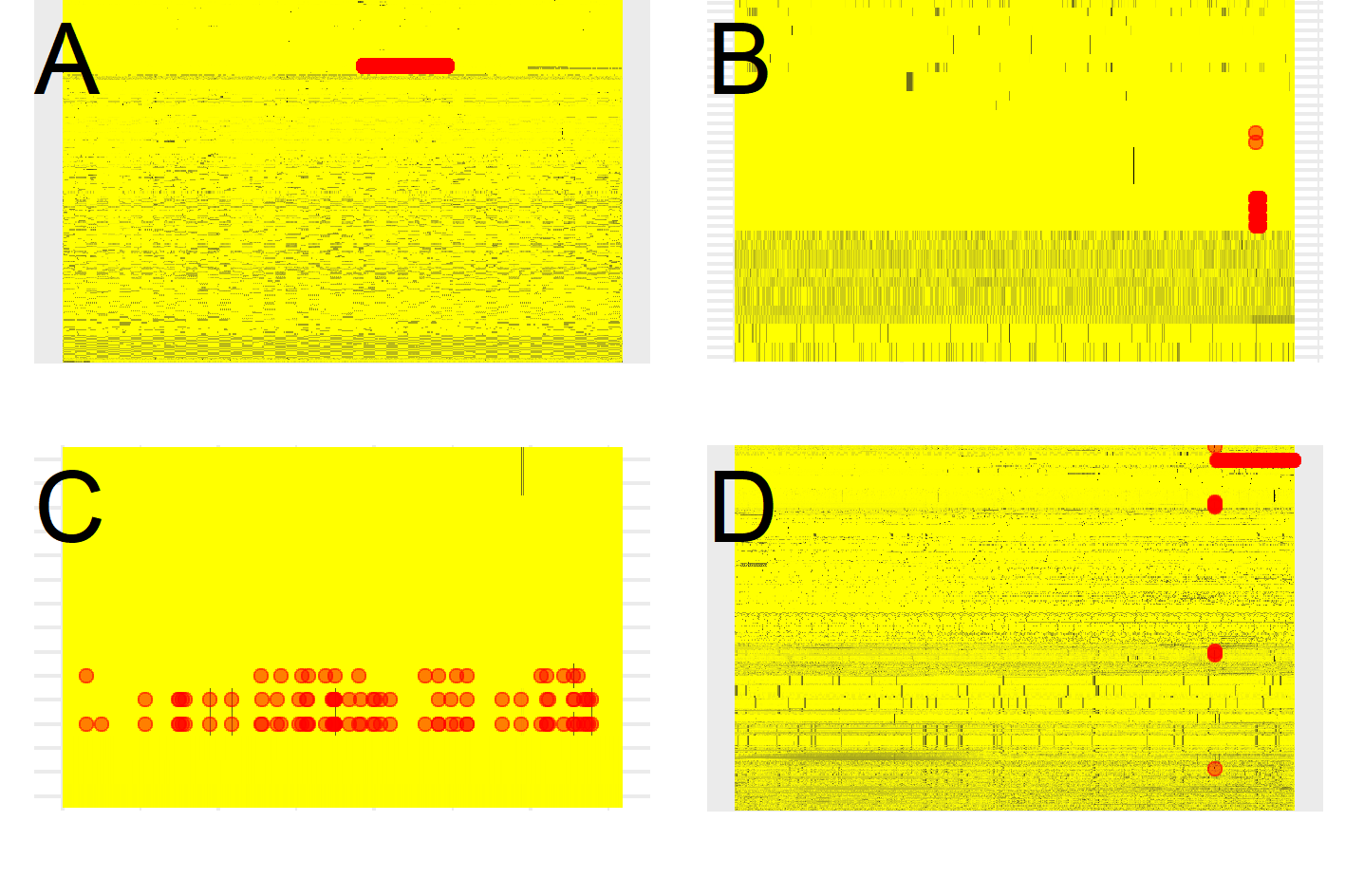}
  \includegraphics[width=0.325\textwidth]{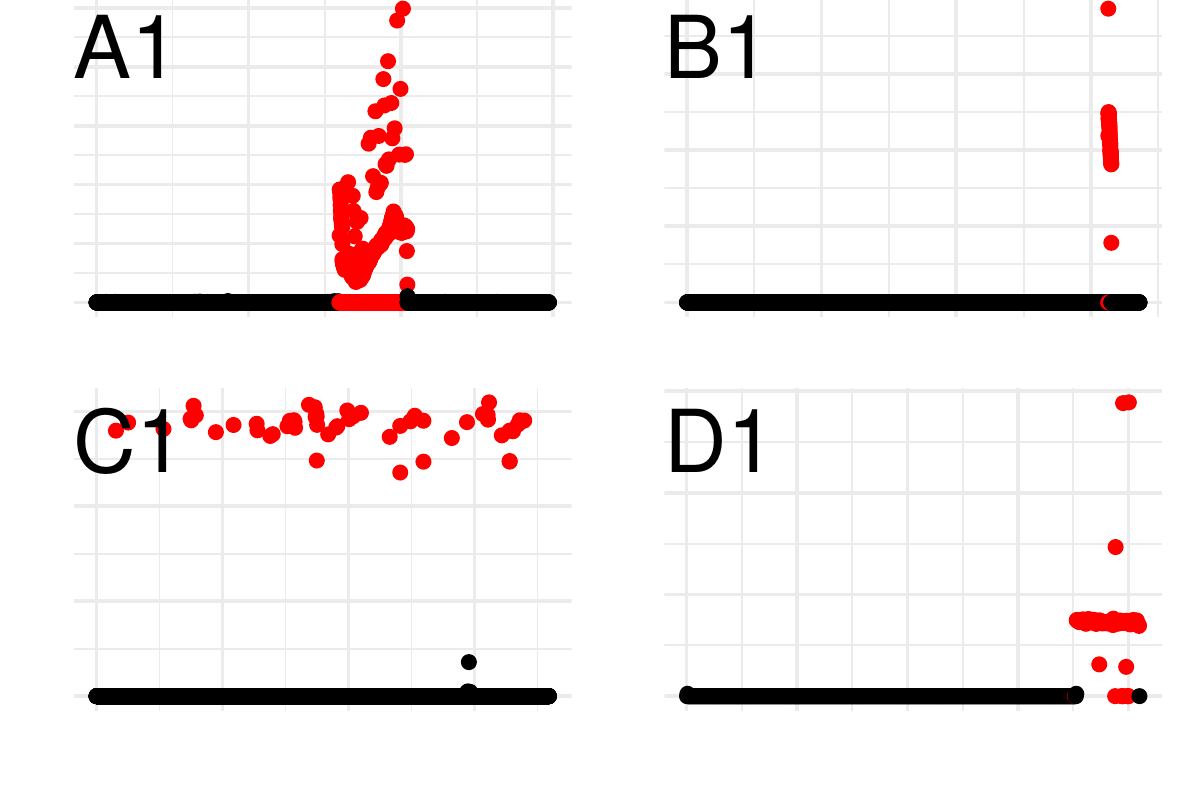}
  \includegraphics[width=0.325\textwidth]{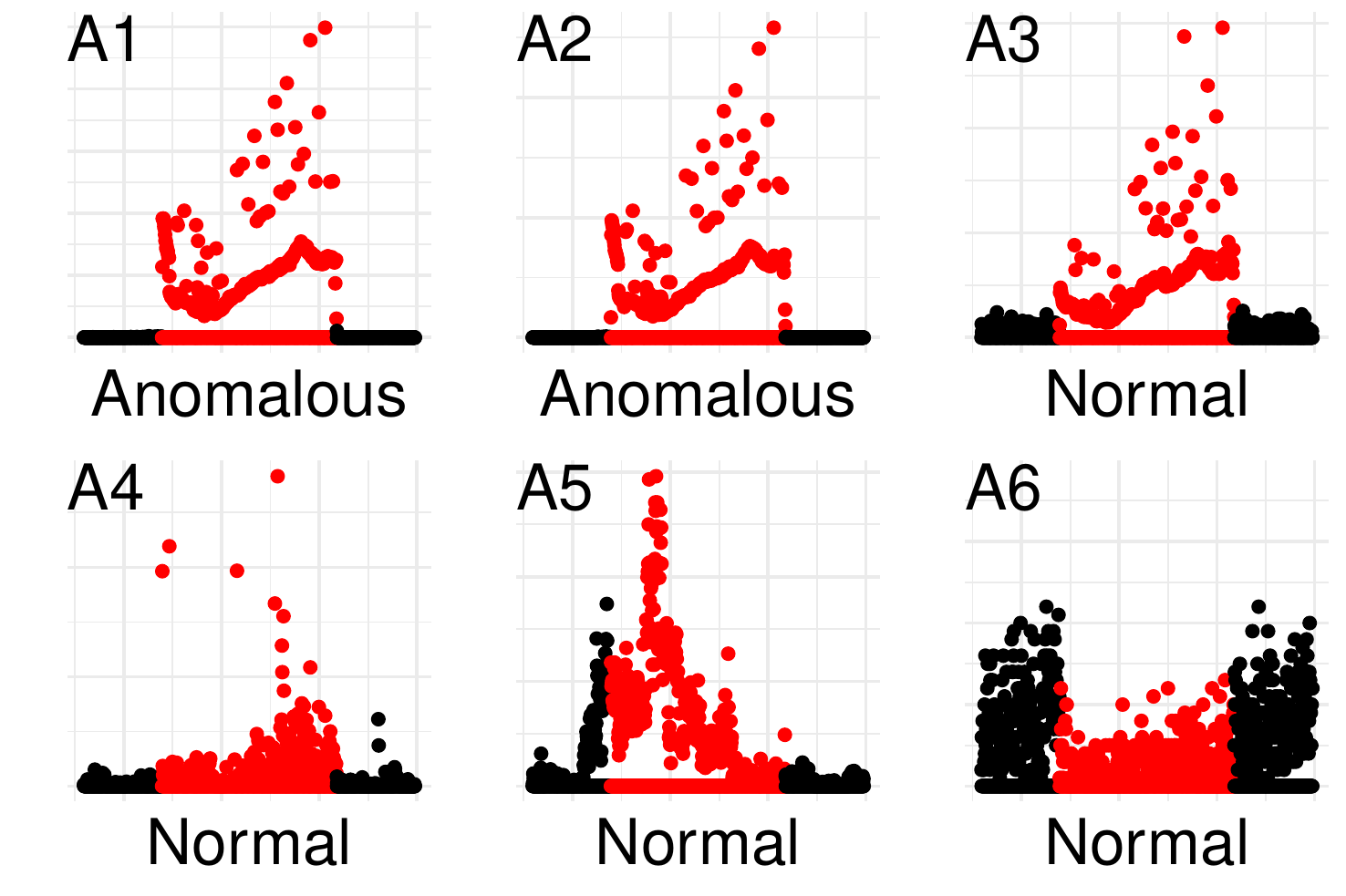} 
(a)\superquad(b)\superquad(c)
  \caption{\textit{Dataset description}. Compact view of anomalies and ground-truth in our datasets: (a) heatmap of KPI over time, with annotated anomalous ground truth (red dots) for four example datasets A-D;  (b) time evolution of the topmost anomalous feature for the example datasets A-D (anomalous period in red); (c) zoom of the topmost 6 anomalous features of dataset A, of which only the first 3 are flagged by the expert.}
  \label{fig:datasets}
\end{figure*}

\subsubsection{Expert knowledge}
A final viewpoint is how to practically leverage Expert Knowledge (EK). Leveraging human expertise in anomaly detection is the focus of \emph{active learning} studies such as  \cite{sota_human_labelless,sota_human_opprentice,sota_human_sfe,sota_human_feedback}, which share similarities with our work.
Authors in~\cite{sota_human_labelless} alleviate the task of generating labels for AD by using a high recall algorithm (producing many false alarm) paired with a human analysis phase, where the experts decide which anomalous ``shapes'' they are interested in finding. In our work we employ the temporal portion of the ground truth information as an AD ``Oracle'' only as a reference performance point to contrast results of AD algorithms.
An Operator's Apprentice is proposed in~\cite{sota_human_opprentice} to study univariate time series: humans are required to label a collection of datasets to train an apprentice (Random Forest), on a set of features extracted from the time series, iterating until a configurable performance threshold is attained. Similarly, \cite{sota_human_sfe} introduces an iterative process (referred to as Sequential Feature Explanation), with the explicit objective of quantifying the number of features needed to be shown to an expert in order for him to be sure about an anomaly's nature. Also ~\cite{sota_human_feedback} includes humans in the AD loop, collecting  experts feedback in an online fashion to improve detection accuracy.

While similar in spirit, several differences arise in the interaction mode (explicit~\cite{sota_human_opprentice,sota_human_sfe,sota_human_feedback} vs absent in HURRA), methodology (iterative~\cite{sota_human_opprentice,sota_human_sfe,sota_human_feedback} vs one-shot; supervised~\cite{sota_human_opprentice} vs unsupervised) and final output (sorted timeslots~\cite{sota_human_feedback} vs sorted features). 
Particularly, by seamlessly exploiting ticket solution while avoiding any explicit additional interaction, HURRA improves accuracy and reduces deployment barriers and training costs (as human interface is unmodified).

\section{Dataset Description}\label{sec:dataset}

 We validate HURRA leveraging a collection of 64 real datasets,  that are succintly characterized in Tab.\ref{table:datasets},  illustrated in Fig.\ref{fig:datasets} and that we describe next considering both from Network domain and Machine learning viewpoints.

\subsubsection{Network expert viewpoint}
Overall, the collected data amounts to 125 days worth of router KPI data, out of which about 4.4\% of timestamps and 5.8\% of KPIs are anomalous: note that this seemingly high fraction of anomalies is intrinsically due to the dataset collection process, i.e., the KPIs data  present in our dataset have been collected precisely since they are anomalous in the first place.

Generally, each dataset is collected locally at a single router of a different ISP, so datasets are completely independent from each other. Additionally, there is a large amount of telemetry data (in the order of 70K KPIs)  collected by routers and available for export. At  the same time, exhaustive collection is prohibitive (due to router CPU resources and O\&M bandwidth usage) so that  datasets at our disposal only include the smaller subset of KPIs that have been manually verified by the expert when solving the case (from 6-1650 KPIs). 

Overall, datasets comprise 8463 KPIs, of which only 1105 are unique: at the same time, datasets are significantly different, since between any pair of datasets there is less than 5\% of common KPIs. This makes the datasets significantly different from 
work that focuses on data-plane indicators of traffic volume~\cite{munz2008itc} and video-streaming~\cite{fiadino2014itc}, since KPIs in our work are very diverse internal router counters, that pertain to both control and data planes. Also, unlike studies where coarse KPIs are constructed over long timescales (up to 24hours in~\cite{morichetta2018lenta}), the routers in our dataset export KPIs at a fast 1 minute timescale.

The expert labels the time at which anomalies occurred, and additionally writes in his report the KPIs that were found to be responsible of the issue. In this process the network expert can decide \emph{not to report} a KPI as anomalous, irrespectively of the time series behavior, as e.g.,  he  (i) deems other KPIs  more relevant, (ii) believes the KPI to be an effect, and not the root cause of the issue (e.g., in one dataset a large number of packets are dropped when their TTL reaches 0, which is however not flagged by the expert as this symptom is caused by a routing loop, correctly flagged by the expert). The availability of this very fine-grained KPI-level data makes the dataset peculiar: as causality is knowingly hard to assess~\cite{pearl,biersak_causality}, and since our ground truth does not explicitly provide causal information, a natural benchmark for HURRA is thus  the extent of agreement between algorithmic and human judgments.

\begin{table}[!t]
\caption{Summary of basic dataset properties.}
\label{table:datasets}
\centering
\begin{tabular}{rrrrr}
\toprule
             & \thead{\#Timeslots \\ (rows)} & \thead{\#KPIs \\ (columns)} & \thead{Anomalous \\ Timeslots\%} & \thead{Anomalous \\ KPIs\%} \\ 
             \midrule
Minimum      & 211            & 6                 & 0.01\%                 & 0.27\%            \\
1st Quartile & 1081           & 30                 & 0.16\%                 & 3.14\%            \\
Median       & 1981           & 86                 & 0.68\%                 & 6.62\%            \\
3rd Quartile & 3131           & 131                & 4.19\%                 & 18.21\%           \\
Maximum      & 11700          & 1650               & 64.01\%                   & 37.50\%  \\ \midrule 
Total        & 181075          & 8463              & 4.44\%                 & 5.80\%\\  \bottomrule                   
\end{tabular}
\end{table}

\usetikzlibrary{arrows,shadows,positioning}

\tikzstyle{int}=[draw, fill=gray!20, minimum height=0.8cm] 
\tikzstyle{init} = [pin edge={to-,thin,black},text width=3cm]
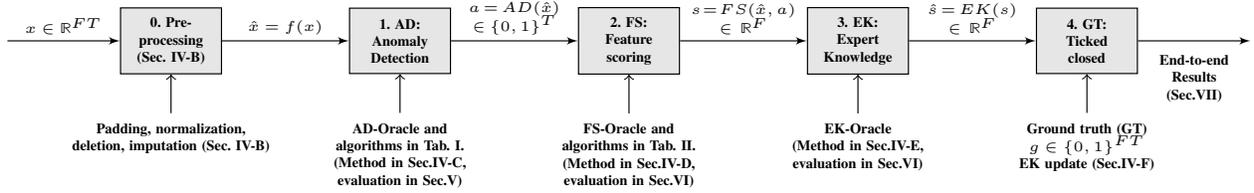
\begin{figure*}[!t]
\resizebox{\textwidth}{!}{%
\begin{tikzpicture}[node distance=2.8cm,text width=1cm,text centered,font=\tiny\bfseries,>=latex']
    \node [int, pin={[init]below: Padding, normalization, deletion, imputation (Sec.~\ref{sec:preprocessing})}] (a) {0.~Pre- processing 
    (Sec.~\ref{sec:preprocessing})};
    \node (start) [left of=a,node distance=2cm, coordinate] {~~~};
    \node [int, pin={[init]below: AD-Oracle and\\algorithms in Tab.~\ref{table:algorithms}.\\ (Method in Sec.\ref{sec:ad}, evaluation in Sec.\ref{sec:results:AD})}] (b) [right of=a] {1.~AD: Anomaly Detection};
    \node [int, pin={[init]below: FS-Oracle and\\algorithms in Tab.~\ref{tabl:sota:fs}.\\ (Method in Sec.\ref{sec:fs}, evaluation in Sec.\ref{sec:results:FS})}] (c) [right of=b] {2.~FS: Feature scoring};    
    \node [int, pin={[init]below: EK-Oracle \\ (Method in Sec.\ref{sec:ek}, evaluation in Sec.\ref{sec:results:FS})}] (d) [right of=c] {3.~EK: Expert Knowledge};    
      \node [int, pin={[init]below: Ground truth (GT)\\$g\!\!\!\!\in\!\!\!\!\{0,1\}^{F T} $\\ EK update (Sec.\ref{sec:update})}] (e) [right of=d] {4.~GT: Ticked closed};    
    \node [coordinate] (end) [right of=e, node distance=2cm]{};
    \path[->] (start) edge node [text width=2.5cm,midway,above=-0.5mm]{$ x\!\!\!\!\in\!\!\!\!\mathbb{R}^{FT}$} (a);
    \path[->] (a) edge node [text width=2.5cm,midway,above=-0.5mm]{$\hat x\!\!\!\!=\!\!\!\!f(x)$} (b);
    \path[->] (b) edge node [text width=2.5cm,midway,above=-0.5mm]{$a\!\!\!\!=\!\!\!\!AD(\hat x)$\\$\in\!\!\!\!\{0,1\}^T$} (c);
    \path[->] (c) edge node [text width=2.5cm,midway,above=-0.5mm]{$s\!\!\!\!=\!\!\!\!FS(\hat x, a)$\\$\in\!\!\!\!\mathbb{R}^F$} (d);
    \draw[->] (d) edge node [text width=2.5cm,midway,above=-0.5mm]{$\hat s\!\!\!\!=\!\!\!\!EK(s)$\\$\in\!\!\!\!\mathbb{R}^F$} (e) ;
    \draw[->] (e) edge node [text width=2.5cm,midway,below=0.5mm]{End-to-end\\ Results\\(Sec.\ref{sec:results:E2E})} (end) ;
\end{tikzpicture}
 } 
\caption{Detailed synoptic of the notation (Sec.~\ref{sec:definition}), workflow (Sec.~\ref{sec:preprocessing}--Sec.~\ref{sec:update}) and evaluation (Sec.~\ref{sec:res:ad}--Sec.~\ref{sec:results:E2E}).}
\label{tikzfigure}
\end{figure*}

\subsubsection{Machine learning expert viewpoint}
Both the collected features (KPIs), as well as the binary ground truth labels can be abstracted as multivariate time series.  On the one hand, since datasets are independently collected and the set of collected features is also largely varying,  supervised techniques such as those exploited in \cite{casas2016machine} are 
not applicable.  On the other hand, the limited number of features reduces the problems tied to the ``curse of dimensionality''. Overall, these observations suggest  unsupervised algorithms to be a good fit.

For the sake of clarity, we  compactly visualize the diversity of our datasets in
Fig.~\ref{fig:datasets}-(a), which depicts 4 samples datasets A--D  as a heatmap (x-axis represents the time, the y-axis a specific feature, and the z-axis heatmap  encodes the standardized feature value, with darker yellow colors for greater deviation from the mean value), with annotated ground truth (red dots). The figure additionally depicts the temporal evolution of  (b) the topmost anomalous feature found by HURRA in each of the 4 samples datasets (normal portion in black,  anomalous portion in red) as well as of (c) the topmost anomalous 6 features for dataset A. Fig.~\ref{fig:datasets}-(a) and (b) clearly illustrate the dataset variability, even from the small sample reported, in terms of the type of outliers present in the data. 

Fig.~\ref{fig:datasets}-(c) further illustrates the semantic of the available ground truth: it can be seen that the top-three features (A1, A2 and A3) are highly correlated and are also widely varying during the anomaly period; yet the subsequent features (A4, A5 and A6) also present signs of temporally altered behavior, yet these time series were investigated by the human operator, but ruled out in the ticket closing the solution -- well highlighting the cause/effect mismatch.

\section{Methodology}\label{sec:methodology}

The main objective of HURRA  is to  sort the KPIs in a multivariate time series in a way  that reduces the time it takes to complete its troubleshooting process. We first outline our full workflow and introduce  some necessary notation  (Sec.\ref{sec:definition}), and next detail the individual system components  (Sec.\ref{sec:ad} through \ref{sec:update}).

\subsection{Definitions}\label{sec:definition}
With reference to the building blocks earlier overviewed in  Fig.~\ref{fig:birdeye}, we now more precisely describe our workflow and notation with the help of Fig.~\ref{tikzfigure}.
We denote the collected multi-dimensional time-series data in as a matrix  $x\in \mathbb{R}^{F T}$, with $F$ the number of existing features and  $T$ the number of timeslots in the dataset. After \circled{0} preprocessing $\hat x = f(x)$ the data becomes an input to the \circled{1} \emph{Anomaly Detection (AD)} function,  whose output is a binary vector $a = AD(x) \in \{0,1\}^T$ indicating if the $i$-th timeslot is anomalous or not. The \circled{2} \emph{Feature Scoring (FS)} function takes both $\hat x$ and $a$ as input, producing a real-valued vector $s = FS(\hat x, a)  \in \mathbb{R}^F$ where $s_j$ represents  an ``anomalous score'' associated to the  $j$-th feature.
An \circled{3} \emph{Expert Knowledge (EK)} function can be used to alter the feature scoring vector $\hat s = EK(s) \in \mathbb{R}^F$ whenever available, else $\hat s = s$.
In this paper, we use \circled{4} \emph{Ground Truth (GT)} labels to assert the performance of our system, as well as to \circled{5} mimic the update of the expert knowledge (more details later). Formally  $g \in \{0,1\}^{F T}$ is a ground-truth matrix, with $g_{jt}=1$  indicating that at the  $t$-th timeslot the $j$-th feature was flagged as anomalous by the expert. It follows that a feature $j$ is  anomalous when $\sum_t g_{jt}>0$, and similarly we can identify anomalous timeslots $a_t$ as the time instants where at least a feature is anomalous, i.e.,  $a_t = \mathbb{1}\left(\sum_j g_{jt}>0\right)$.

\subsection{Data preprocessing}\label{sec:preprocessing}
As our methodology is designed with an inherently practical goal,
we cannot assume clean input data. As such, we need to ensure quality of the 
multi-dimensional timeseries $x$, by performing the usual data sanitization steps of:
(i)  time series alignment and padding, (ii) removal of constant features, or features with excessive fraction ($>50\%$) missing samples,
(iii) sample-and-hold imputation of remaining missing samples, (iv)  standard normalization of individual KPIs $\hat x= (x-\mathbb{E}[x])/\sigma_X$, which is necessary only in the case of distance/density based algorithms.

At the end of this preprocessing pipeline, the streaming data $\hat x$ is composed of equally spaced time-series, with no missing samples or constant features, ready to be processd by  the Anomaly Detection module.

\subsection{Anomaly Detection (AD)}\label{sec:ad}
As overviewed in Sec.\ref{sec:related:ad}, the literature is ripe with AD methods: hence, the main focus of our system is not to propose a new, better, algorithm, but rather to exploit the best in class, and possibly combining multiple algorithms.
To do so, we include representatives of each family of algorithms in our tests, ensuring also the presence of batch and streaming ones. Specifically, 
\begin{itemize}
    \item  for proximity-based algorithms we use DBSCAN, LOF (batch) and OADDS (streaming); 
    \item for tree-based algorithms we use IF and RHF (batch), RRCF and HST  (streaming);
    \item for subspace-based algorithms we use LODA and xStream (streaming).
\end{itemize}

While details of each algorithm settings are deferred to the performance evaluation section Sec.\ref{sec:res:ad}, it is worth stressing that for each algorithm, we further consider two scenarios of application:
\begin{itemize}
\item \emph{Practical Lower-Bound (LB):} the case of a single hyperparametrization for all datasets 
\item \emph{Ideal Upper-Bound (UB):}  the case of multiple  hyperparametrizations, fit for each datasets by means of a standard grid search procedure
\end{itemize}
\noindent LB and UB are specifically intended to  gather  a conservative and optimistic performance assessment of the AD module, respectively. The lower-bound is indeed representative of a simple practical system, where AD settings are determined on a set of use-cases at time $0$, that are never changed later on. The upper-bound is instead representative of a system where an hypothetical automated procedure is able to optimally select the AD algorithm parameters, which in our case is done a posteriori using the the ground truth labels $a_t$. 

Additionally, as a reference we further consider two  ideal  benchmarks:
\begin{itemize}
\item \emph{Ideal Ensemble:} selects the AD results given by the best performing algorithm, using  the ground truth labels $a_t$ to perform the algorithm selection.
\item \emph{AD Oracle:} directly uses the ground truth labels to identify anomalous timeslots $a_t$, with perfect agreement with the expert.
\end{itemize}
\noindent Despite the above benchmarks are clearly impractical,  the AD Oracle is important as it allows us to decouple the evaluation of the AD and FS building blocks, by examining impact of FS under ideal solution of the AD problem. 
Moreover,  contrasting the Ideal ensemble vs AD Oracle performance allow to understand if a broader algorithm selection is needed to approach the ideal solution.  

\subsection{Feature Scoring (FS)}\label{sec:fs}
The purpose of the FS step is to present human operators with features that are relevant for the solution of the troubleshooting. From a purely data-driven viewpoint, the intuition is that network experts would want to look first at  KPIs  exhibiting drastic change when anomaly happens.  We propose 2 novel practical FS strategies and compare them against  a set of 8 reference strategies. 

\subsubsection{Proposed FS strategies}
Our contribution here is to aim for non-parametric FS functions, to avoid introducing further arbitrary  hyperparameters beyond those of AD algorithms.

\paragraph{Average-based FS (FSa)}
In particular, we define a first $FSa$ function sorting KPIs by their average difference between anomalous and normal times. 
Formally, given $\sum_t a_t$ the number of anomalous timeslots found by the AD algorithm (so that $T-\sum_t a_{t}$ is the number of normal timeslots), the $j$-th feature anomalous score is defined as:
\begin{equation}\label{eq:FSa}
s^{FSa}_{j} = \left| \frac{\sum_t a_{t} \hat x_{jt}}{\sum_t a_{t}}  -  \frac{\sum_t  (1-a_{t})\hat x_{jt}}{T-\sum_t a_{t}}   \right| 
\end{equation}
It is useful to recall that, since $\hat x$ features are normalized, the
magnitude of the scores $s^{FSa}_{j}$ returned by $FSa$  is directly comparable.

\paragraph{Rank-based FS (FSr)}
We additionally define $FSr$  to measure the difference in the ranking of the $j$-th feature during anomalous vs normal times. Letting normal and anomalous ranks $r^-_j$ and $r^+_j$ respectively:
\begin{eqnarray}
r^+_j &=& \textrm{rank}\left(j : \textstyle\sum_t  a_{t}\hat x_{jt}/\textstyle\sum_t a_{t} \right) \\
r^-_j &=& \textrm{rank}\left(j : \textstyle\sum_t  (1-a_{t})\hat x_{jt}/(T-\textstyle\sum_t a_{t})    \right)
\end{eqnarray}
\noindent the rank-based score becomes:
\begin{equation}\label{eq:FSr}
s^{FSr}_j =  \left| r^+_j -  r^-_j \right|
\end{equation}
Intuitively, whereas $FSa$ compares variations of normal vs anomalous features values in the normalized \emph{feature value} domain, $FSr$ compares the relative impact of such  changes in the \emph{features order} domain -- somewhat analogous to Pearson vs Spearman correlation coefficients.

\subsubsection{Reference FS strategies}
We compare our novel FS functions with a set of benchmarks of increasing complexity.

\paragraph{Random order} a na\"ive strategy where KPIs are sorted at random, representing a FS performance lower bound.

\paragraph{Alphabetical order} the standard approach, where KPIs are presented in alphabetical order (e.g., following a structured YANG-tree model), that is convenient to browse for human operators but is not otherwise expected to provide significant benefits over random ordering. 

\paragraph{Normal ranking}
following the methodology proposed in \cite{sota_other_adele}, we rank each KPI $j$ by fitting a  Normal distribution $\mathcal{N}(\mu_j,\sigma_j)$ using the samples not labeled as anomalous by the AD algorithm. Then, each anomalous sample $x_{jt}$ is assigned a score using the previously fitted normal  Cumulative Distribution Function (CDF), as follows:

\begin{equation}
    s_{jt}=2\left|\frac{1}{2} - \frac{1}{\sigma_{j}\sqrt{2 \pi}}e^{-(x_{jt}-\mu_{j})/2\sigma_{j}}\right|
\end{equation}
\noindent and the final score of KPI $j$ is calculated as the average of the individual scores of anomalous samples respective to it:
\begin{equation}
    s^{ND}_j=\frac{1}{\sum_t a_t}\sum_t s_{jt}
\end{equation}

\paragraph{Lognormal ranking}
in practice, we do not expect network KPIs to be well modeled as a mono-modal normal distribution. As such,  
we account for non symmetric and skewed data by further introducing a log-normal scoring function as follows:

\begin{equation}
    s^{NDLog}_j = \frac{1}{\sum_t a_t}\sum_i -\log{(1-s_{jt})}
\end{equation}

\paragraph{Ideal predictive FS}
We finally consider a family of four predictive policies inspired by PROTEUS~\cite{PROTEUS}, that first builds a supervised model using the AD algorithm result as  labels and   then exploits the feature importance for classification task (i.e., which features have the highest predictive power to classify the anomaly) as feature ranking scores.

Specifically, we test four different policies, based on  two classifiers: (i) Random Forest (RF)~\cite{randomforests}, the most popular tree based ensemble classification method, using the impurity criterion for the extraction of feature importance $s^{RF}$; and (ii) Elastic Net Logistic Regression (ENLR)~\cite{elasticnet},
using the regression coefficients as feature importance values. ENLR allows for feature selection through the linear weighting of L1 and L2 regularization based on a parameter $\alpha$:  we get $s^{Lasso}$ for  Lasso when  $\alpha = 1$,   $s^{RR}$ for Ridge Regression (RR) when $\alpha = 0$ and an equal combination of both $s^{EN}$ for $\alpha = 0.5$. In all cases, predictive scores $s^{RF}, s^{Lasso}, s^{RR}, s^{EN}$ are used in the same way as for all previous policies, i.e., to rank features by their importance in explaining the anomalous event.

\subsection{Expert Knowledge (EK)}\label{sec:ek}

By leveraging previously solved cases, one can easily build an EK-base.
Intuitively, if for cases where KPIs $A$, $B$ and $C$ have similar behavior (and thus similar FS scores), only KPI $B$ is labeled as anomalous, 
we can 
assume that KPI $B$ is the only one that experts are interested in seeing to correctly diagnose the case: a ranking more useful for the human operator could be obtained by altering the results of the FS block, e.g. reducing $s_A$ and/or increasing of $s_B$. 
Tracking for each KPI\footnote{Clearly a given feature named $J$ does  not maintain  the same index $j$ across different datasets; however, for the sake of simplicity we prefer to abuse the  notation and  confuse the feature name and index name in this sub-section} $j$ the number $n_j$ of troubleshooting cases where $j$ was observed, as well as the number of times $n^+_j$ it was flagged as anomalous, one can gather its anomaly rate:
\begin{equation}
K^+_j=n^+_j/n_j    \label{eq:splus} 
\end{equation}
Additionally,  EK  can also track, over all datasets, the  number $n^-_j$ of cases where feature named $j$ is \emph{not} flagged by the expert ($g_{jt}=0, \forall t$), despite its score during the anomalous period being larger than the minimum score among the set $\mathcal{A}$ of other features flagged as anomalous:
\begin{equation}
s_j > \min\limits_{k\in \mathcal{A}} s_k,  \quad  \mathcal{A} = \left\{k : \textstyle\sum_t g_{kt}>0 \right\}  \label{eq:sminus}
\end{equation}
Explicitly considering the fact that the expert actually ``ignored'' feature $j$ at a rate  $K^-_j=n^-_j/n_j$ can be helpful in reducing the attention on features that are considered to be less important by the expert (e.g., as they may be effects rather than causes).
The knowledge base can be queried to alter the scores of the FS step as in:
\begin{equation}
\hat s_j= s_j (1+\gamma^+ K^+_j -\gamma^- K^-_j)
\label{eq:fs}
\end{equation} 
by positively biasing ($K^+_j$) scores of KPIs that were found by experts to be the culprit in previous cases, and negatively biasing ($K^-_j$) those that were not flagged by the expert despite having a large anomalous score. In (\ref{eq:fs}) the free parameters $\gamma^+,\gamma^-\in\mathbb{R}^+$  allow to give more importance to past decisions ($\gamma >  1$) or to the natural scoring emerging solely from the current data ($\gamma \rightarrow 0$). 

Interestingly, the frequentist approach can assist in the resolution of common problems (where the $K^+_j$ rate is high and the number of observation $n_j$ is equally large), whereas in uncommon situations (where a new KPI is the culprit), it would still be possible to ignore EK suggestions by reducing $\gamma \rightarrow 0$ (requiring a slight yet intuitive addition to the human interface design). De facto, the use of EK alters FS scores in a semi-supervised manner (and so is able to improve on known KPIs), whereas it leaves the AD block unsupervised (and so able to operate on previously unseed KPIs).

\subsection{Update of Expert Knowledge Base}\label{sec:update}
We remark that EK management is particularly lean in HURRA. In case of cold-start (i.e., no previous EK base),  $K^+_j=0$ so that $\hat s_j = s_j$ by definition. Additionally, the system can operate in an incremental manner and is capable of learning over time without explicit human intervention:  the information is extracted from the tickets by simple update rules (\ref{eq:splus})--(\ref{eq:sminus}).
Finally, the EK update mechanism supports transfer learning, as it is trivial to ``merge'' knowledge bases coming from multiple ISPs, by e.g., simple weighted average of $K^+_j$ and $K^-_j$ rates. As a matter of fact, our datasets already aggregates multiple ISPs deployment, which can therefore prove (or invalidate) the transferability of our proposed semi-supervised mechanism.

In this work, we simulate the EK update process by performing \emph{leave-one-out} cross-validation: i.e., upon analysis of a given dataset, 
we only consider knowledge coming from the other datasets, which prevents overfit. 
Also, in reason of the significant heterogeneity (very few KPIs in common across datasets), and the relative low number of datasets, the evaluation in this paper constitute a  stress-test w.r.t. the expected performance of a massively deployed EK system having collected thousands of instances.

\section{Anomaly Detection}\label{sec:results:AD}

\begin{table}[!t]
\caption{Hyperparameter search space for AD algorithms.}
\label{tab:params}
\begin{tabular}{ll}
\toprule
Algorithm (N@T)                                                   & Hyperparameters$^1$                                                                                                                                                                                                                                                                               \\ \midrule 
\begin{tabular}[c]{@{}l@{}}DBSCAN\\ (260@8hr)\end{tabular}           & \begin{tabular}[c]{@{}l@{}}$\epsilon \in \{1,2,3,\dots, \textbf{13}, \dots, 20\}$\\ $\textrm{Leaf size} = 30$\\ $\textrm{Minimum samples} \in \{\textbf{2},5,10,20,40,\dots,200\}$\end{tabular}                                                                                                                                 \\ \hline
\begin{tabular}[c]{@{}l@{}}HST\\ (12@4hr)\end{tabular}               & \begin{tabular}[c]{@{}l@{}}$\psi \in \{1\%, 3\%, 10\%, \textbf{30\%}\}$\\ $\textrm{Number of trees} \in \{100, 200, \textbf{300}\}$\\ $h = 10$\end{tabular}                                                                                                                                                         \\ \hline
\begin{tabular}[c]{@{}l@{}}IF\\ (125@1hr)\end{tabular} & \begin{tabular}[c]{@{}l@{}}$\textrm{Number of trees} \in \{10, 50, 100, \textbf{200}, 300\}$\\ $\textrm{Samples per tree} \in \{10\%, 25\%, 50\%, \textbf{75\%}, 100\%\}$\\ $\textrm{Features per tree} \in \{10\%, 25\%, \textbf{50\%}, 75\%, 100\%\}$\end{tabular}                                                           \\ \hline
\begin{tabular}[c]{@{}l@{}}LODA$^2$\\ (4@2hr)\end{tabular}               & $\textrm{Window size} \in \{1\%, 3\%, 10\%, \textbf{30\%}\}$                                                                                                                                                                                                                                             \\ \hline
\begin{tabular}[c]{@{}l@{}}LOF\\ (243@8hr)\end{tabular}              & \begin{tabular}[c]{@{}l@{}}$\textrm{Neighbors} \in \{1,5,25,50,\textbf{75},100,150,200,300\}$\\ $\textrm{Leaf size} \in \{1,5,\textbf{25},50,75,100,150,200,300\}$\\ $p \in \{\textbf{1}, 1.5, 2\}$\end{tabular}                                                                                                               \\ \hline
\begin{tabular}[c]{@{}l@{}}MCOD\\ (324@16hr)\end{tabular}              & \begin{tabular}[c]{@{}l@{}}$\textrm{Radius} \in \{0.01,0.03,0.1,0.3,1,3,\textbf{10},30,100\}$\\ $\textrm{Neighbors} \in \{\textbf{3},10,30,100\}$\\ $\textrm{Window size} \in \{1\%, 10\%, \textbf{30\%}\}$ \\ $\textrm{Initial samples} \in \{1\%, \textbf{10\%}, 30\%\}$\end{tabular}                                                                                                               \\ \hline
\begin{tabular}[c]{@{}l@{}}OADDS$^3$\\ (240@2hr)\end{tabular}            & \begin{tabular}[c]{@{}l@{}}$\textrm{Sample skip} \in \{3, 10, 30, \textbf{100}\}$\\ $\lambda \in \{0.10, 0.13, \textbf{0.16}, 0.19\}$\\   $\beta \in \{0.003, 0.01, 0.03, 0.1, 0.2, \textbf{0.3}\}$\\ $tp \in \{30, \textbf{100}\}$\end{tabular} \\ \hline
\begin{tabular}[c]{@{}l@{}}RHF\\ (4@1hr)\end{tabular}                & \begin{tabular}[c]{@{}l@{}}$\textrm{Number of trees} = 100$\\ $\textrm{Maximum height} \in \{\textbf{5},6\}$\\ $\textrm{Check duplicates} \in \{\textbf{\textrm{yes}}, \textrm{no}\}$\end{tabular}                                                                                                                  \\ \hline
\begin{tabular}[c]{@{}l@{}}RRCF$^4$\\ (1@10hr)\end{tabular}               & \begin{tabular}[c]{@{}l@{}}$\textrm{Number of trees} = 10$\\ $\textrm{Shingle size} =10\%$\\ $\textrm{Tree size} = 10\%$\end{tabular}                                                                                                                                                        \\ \hline
\begin{tabular}[c]{@{}l@{}}xStream$^5$\\ (64@16hr)\end{tabular}          & \begin{tabular}[c]{@{}l@{}}$k \in \{50, 100, \textbf{200}, 300\}$\\ $c \in \{50, \textbf{100}, 200, 300\}$\\ $d \in \{10\}$\\ $\textrm{Initial samples} \in \{1\%, 3\%, 10\%, \textbf{30\%}\}$\end{tabular}                                                                                                       \\ \bottomrule
\multicolumn{2}{l}{$^1$Single settings of practical Lower Bound (LB) scenario in \textbf{boldface}}\\
\multicolumn{2}{l}{$^2$LODA: sparsity=true, histogram=continuous, maximum bins=auto}\\ 
\multicolumn{2}{l}{$^3$OADDS: $\epsilon$=\textrm{dynamic}, $\mu$=\textrm{automatic} }\\
\multicolumn{2}{l}{$^4$RRCF: number of tests limited due to computational complexity}\\
\multicolumn{2}{l}{$^5$xStream: rowstream=true, number of windows=1, scoring=auto}\\

\end{tabular}
\end{table}

We start by assessing performance of the AD algorithms  listed in Tab.\ref{tab:params} with a two-fold goal:
first, we aim at contrasting performance of AD in ideal academic settings (i.e., 
where a data scientist has infinite time/CPU budget, and has ground truth at its disposal to fine-tune the AD module performance) vs the expected AD performance in practical settings (i.e., where the choice of the data scientist are cast in stone in the system, and no human operator will alter the settings after deployment). Second, we aim at narrowing down the AD algorithms from our initial large choice, selecting the best candidates for deployment.
We now define the metrics (Sec.\ref{sec:ad:metrics}), that we use to broadly illustrate the performance of AD algorithms (Sec.\ref{sec:ad:glance}) before diving  deeper  into the performance of each algorithm  (Sec.\ref{sec:ad:select}).

\subsection{Performance metrics}\label{sec:ad:metrics}

AD algorithms return anomalous scores associated to a timestamp, which  can be binary or real-valued. To assess the ability of the AD algorithm to detect anomalous timeslots with respect to the ground truth, the generally used procedure is to  evaluate if the scores associated to samples labeled as anomalous in the ground truth are higher than scores associated to non-anomalous samples. 

More precisely, we evaluate AD performance using standard metrics that can be extracted from the classic confusion matrix in information retrieval.
Let us denote  with  \emph{True Positives} ($TP$) the number of correctly classified anomalous sample (i.e., a real anomaly is detected), with \emph{False Positives} ($FP$)  the number of incorrectly classified anomalous samples (i.e., a false alarm is triggered on a normal sample) and  with  \emph{False Negatives} ($FN$), the number of incorrectly classified normal points (i.e., a real anomaly is unnoticed). \emph{Precision} and \emph{recall} are then defined as:
\begin{align}
\textrm{Precision} &=\frac{TP}{TP+FP}\\ 
\textrm{Recall} & = \frac{TP}{TP+FN}
\end{align}
\noindent in a nutshell, precision is the fraction of correctly raised alarms (including false alarms at the denominator) and relates to the trust the operator can put in the alarms.  Recall is the fraction of detected alarms (including the missed  ones at the denominator) and relates to coverage capability of the method. 

Intuitively, precision and recall tradeoff: to most suitable metric to  appreciate this tradeoff is the \emph{Area Under the Precision-Recall Curve (Pr-Rec AUC)}. The Pr-Rec curve is built by setting a threshold from the minimum score to the maximum one and evaluating at each point the precision and recall metrics, then the Pr-Rec AUC is measured.

One important remark is worth stressing. Typically, in classification problems one would use an analogous tool, namely the \emph{Receiver Operator Curve} (ROC) and metric, namely the ROC AUC.  However, the ROC curve 
exposes the tradeoff between \emph{False Positive Rate} ($FPR=\frac{FP}{FP+TN}$) 
and the \emph{True Positive Rate} ($TPR=\frac{TP}{TP+FN}$)  which is not at all suitable for \emph{unbalanced} datasets. Clearly, anomaly being a rare event, the number of true normal samples  (or more precisely, $FP+TN$) is significantly larger (one to two orders of magnitude) that the number or truly anomalous samples ($TP+FN$): intuitively,  the ROC AUC curve does not allow to detect algorithms with poor recall. More formally, while the ROC AUC has a constant range from 0.5 (equal to a random detector) to 1 (a perfect one), Pr-Rec AUC's minimum value  depends on the dataset it is being calculated on, and specifically relates to the proportion of the minority class to detect ($\textrm{Pr-Rec AUC}_{min} = \frac{N}{N+P}$, with $N$ being the number of members of the minority class and $P$ being the number of members of the majority class). This makes Pr-Rec AUC more suitable than ROC AUC to assess anomaly detection capabilities.

\begin{figure}
    \centering
    \includegraphics[width = \columnwidth]{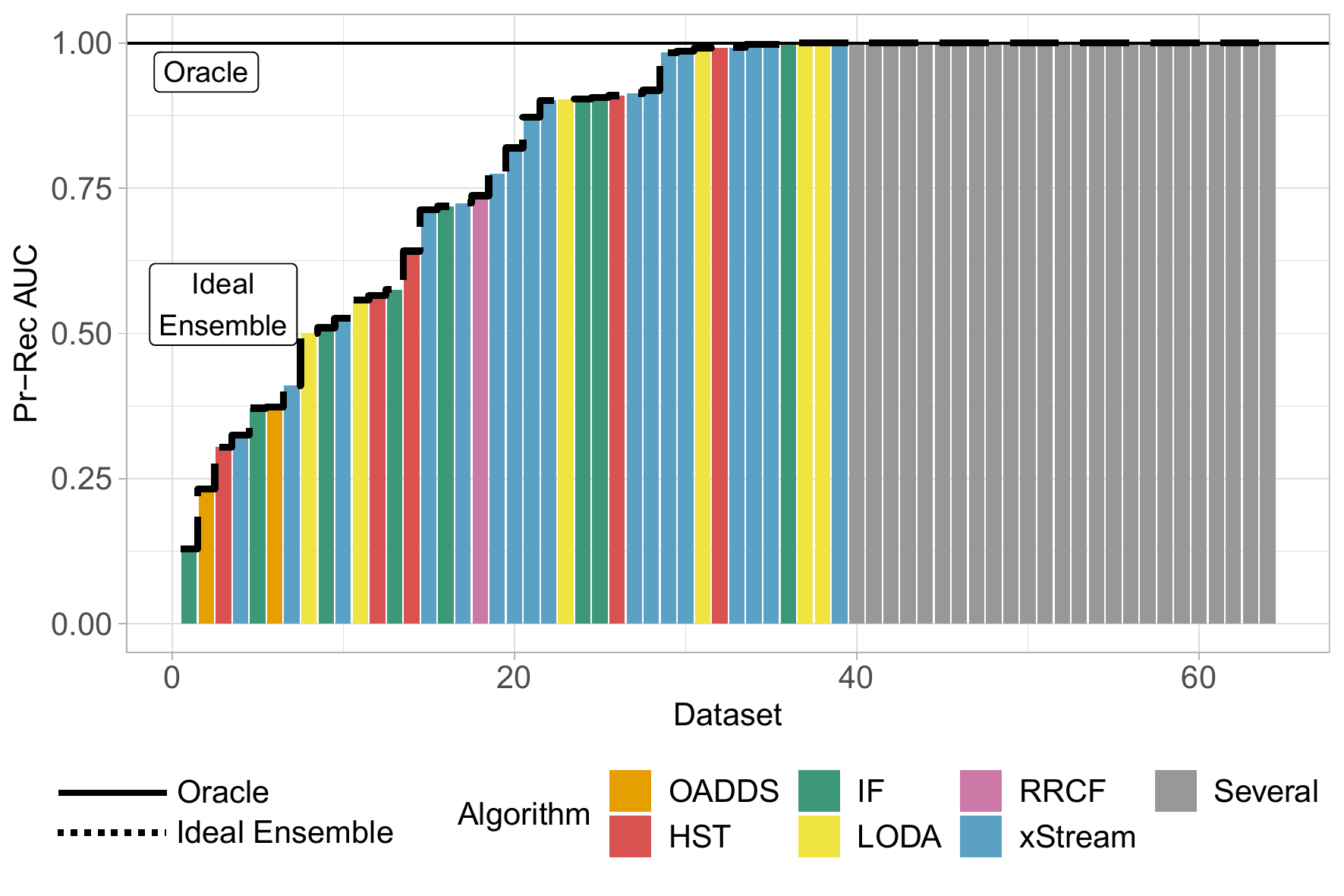}
    \caption{\textit{Results at a glance}:  Pr-Rec AUC performance of the 
    AD-Oracle  (black solid line), of the Ideal ensemble (black dashed line) and the algorithm selected by the ideal ensemble (colored bars).}
    \label{fig:ad_1_best}
\end{figure}

\subsection{Results at a glance}\label{sec:ad:glance}

\subsubsection{Hyperparameters}

\begin{figure*}[!t]
\centering
    \includegraphics[width=0.9\columnwidth]{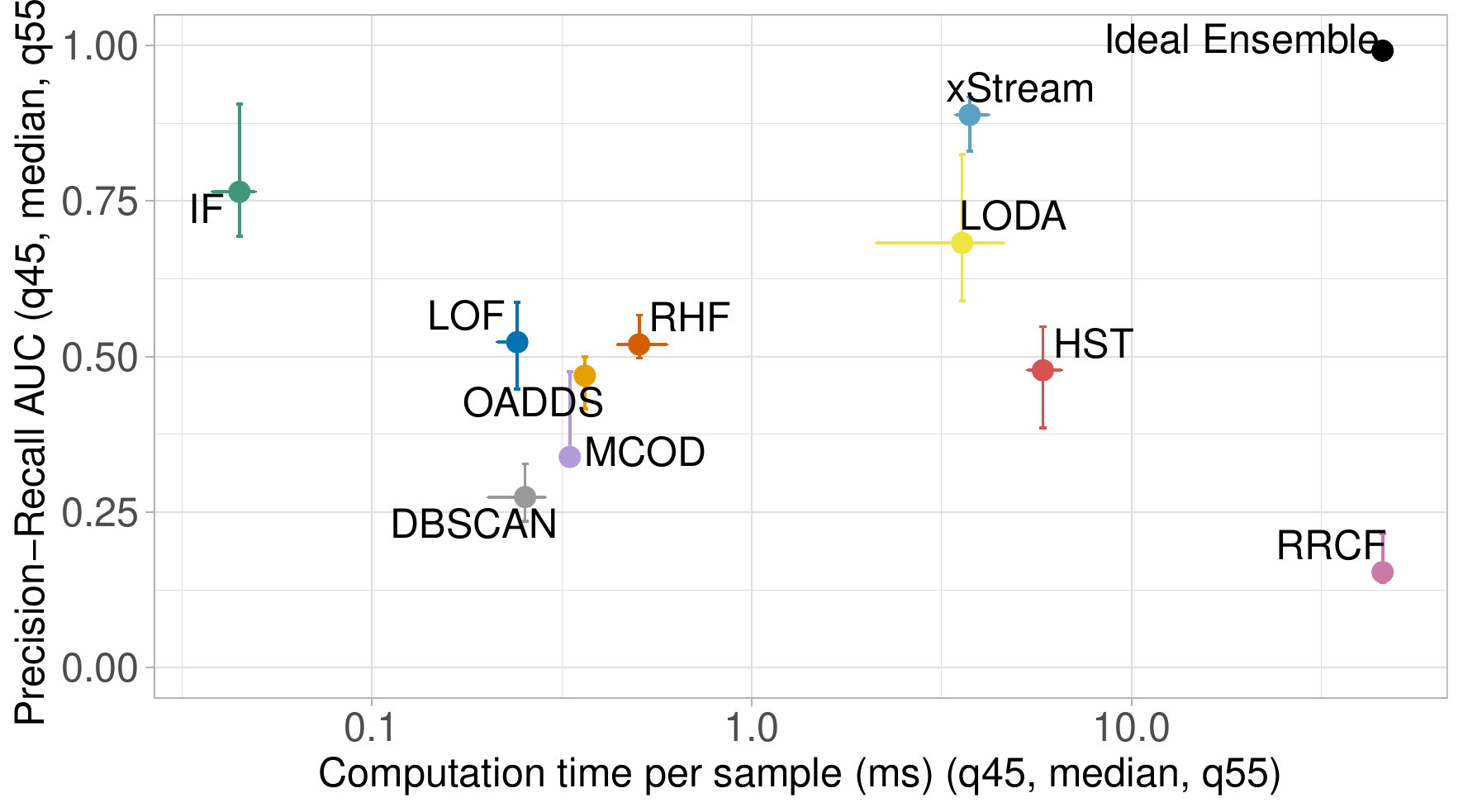}~~~~~ 
   \includegraphics[width=0.9\columnwidth]{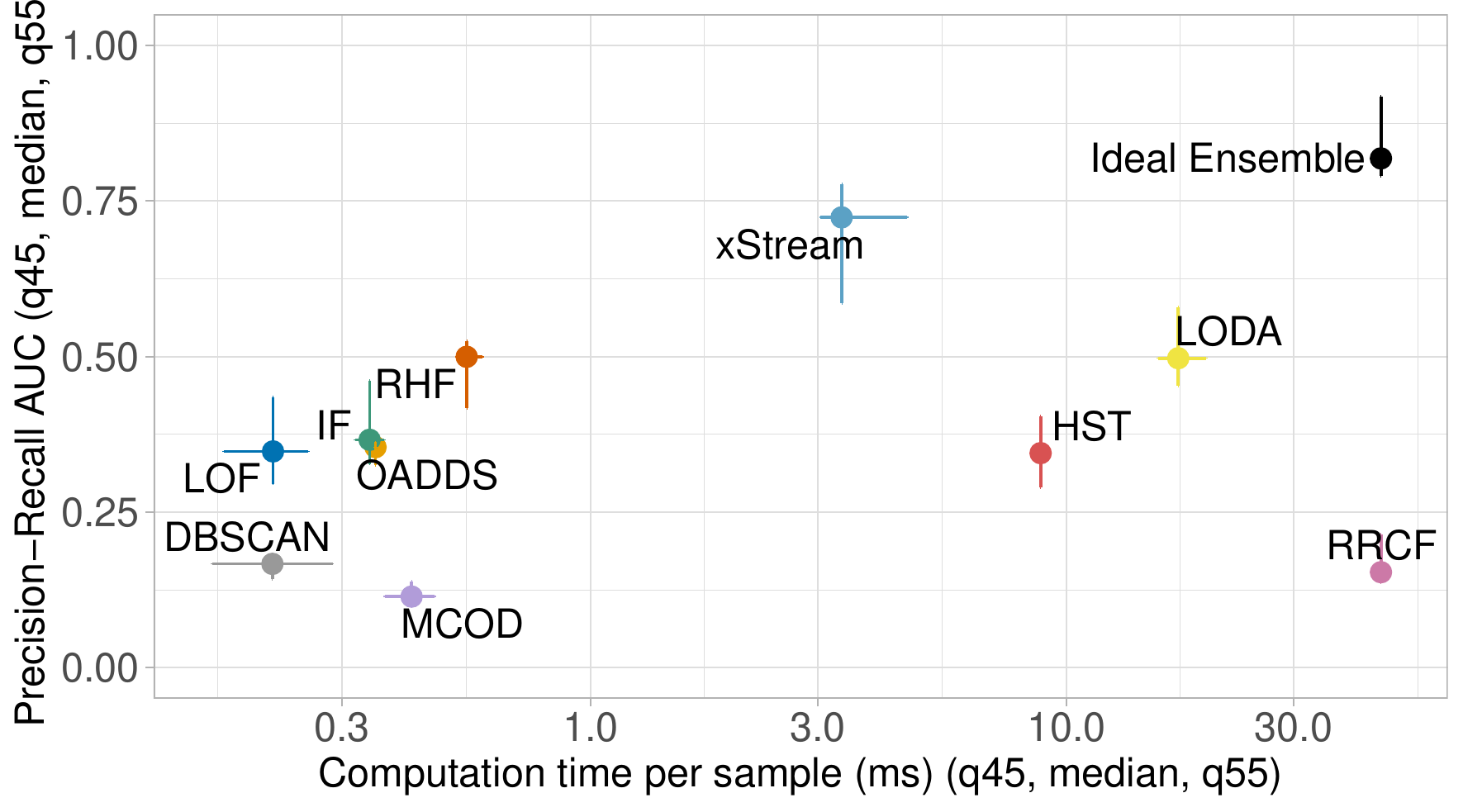}   
   (a)$\qquad\qquad\qquad\qquad\qquad\qquad\qquad\qquad\qquad\qquad\qquad\qquad$(b)
    \caption{\textit{Selection of AD algorithm}:  scatter plot of median Pr-Rec AUC vs median computation time across 64 datasets for each algorithm, including Ideal ensemble as a reference. 
    (a) Ideal upper bound (multiple hyperparametrizations)  and (b) Practical lower bound (single hyperparametrization) scenarios.}
    \label{fig:ad_2_best}
\end{figure*}

\begin{figure*}[!t]
\centering
\includegraphics[width=\columnwidth]{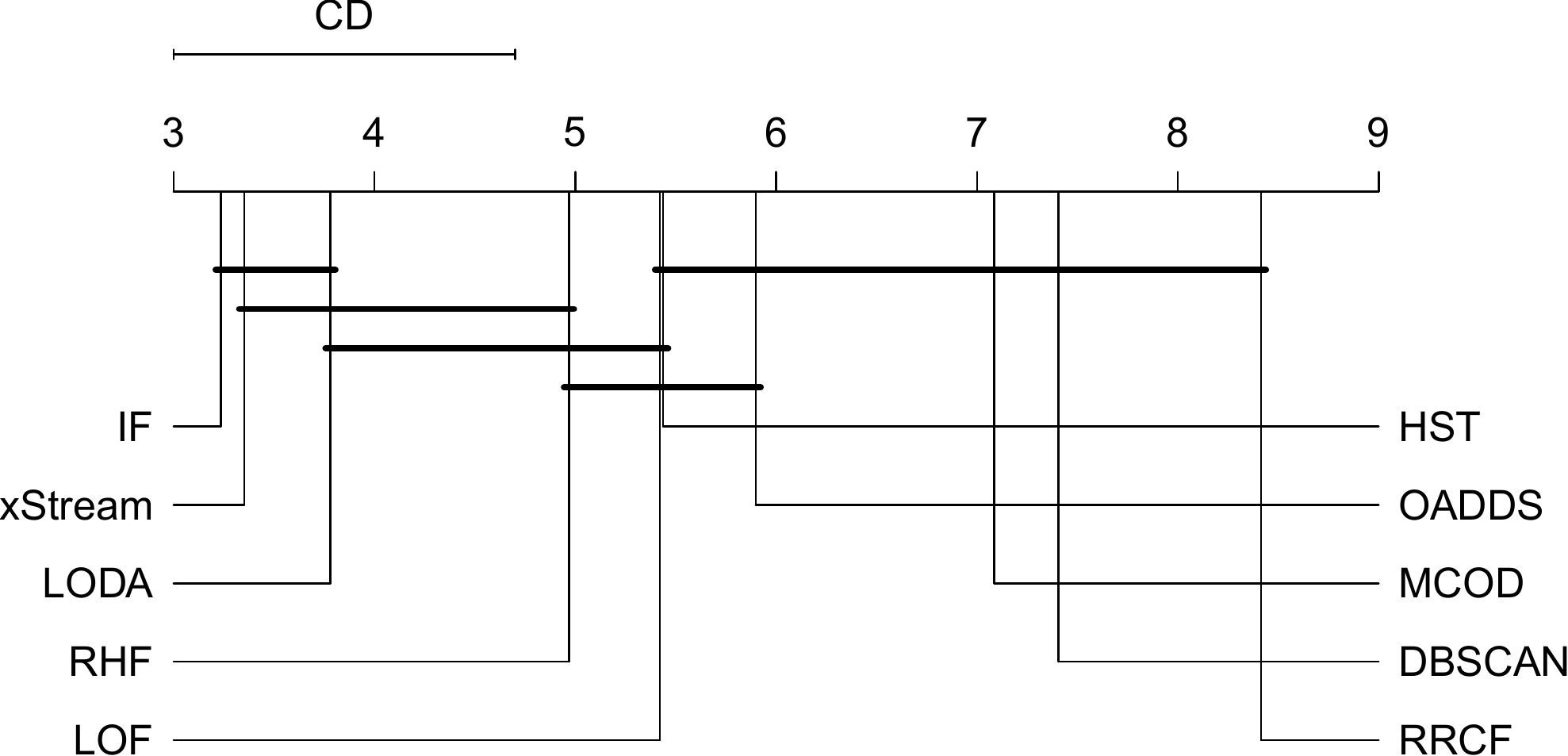}
    \includegraphics[width = \columnwidth]{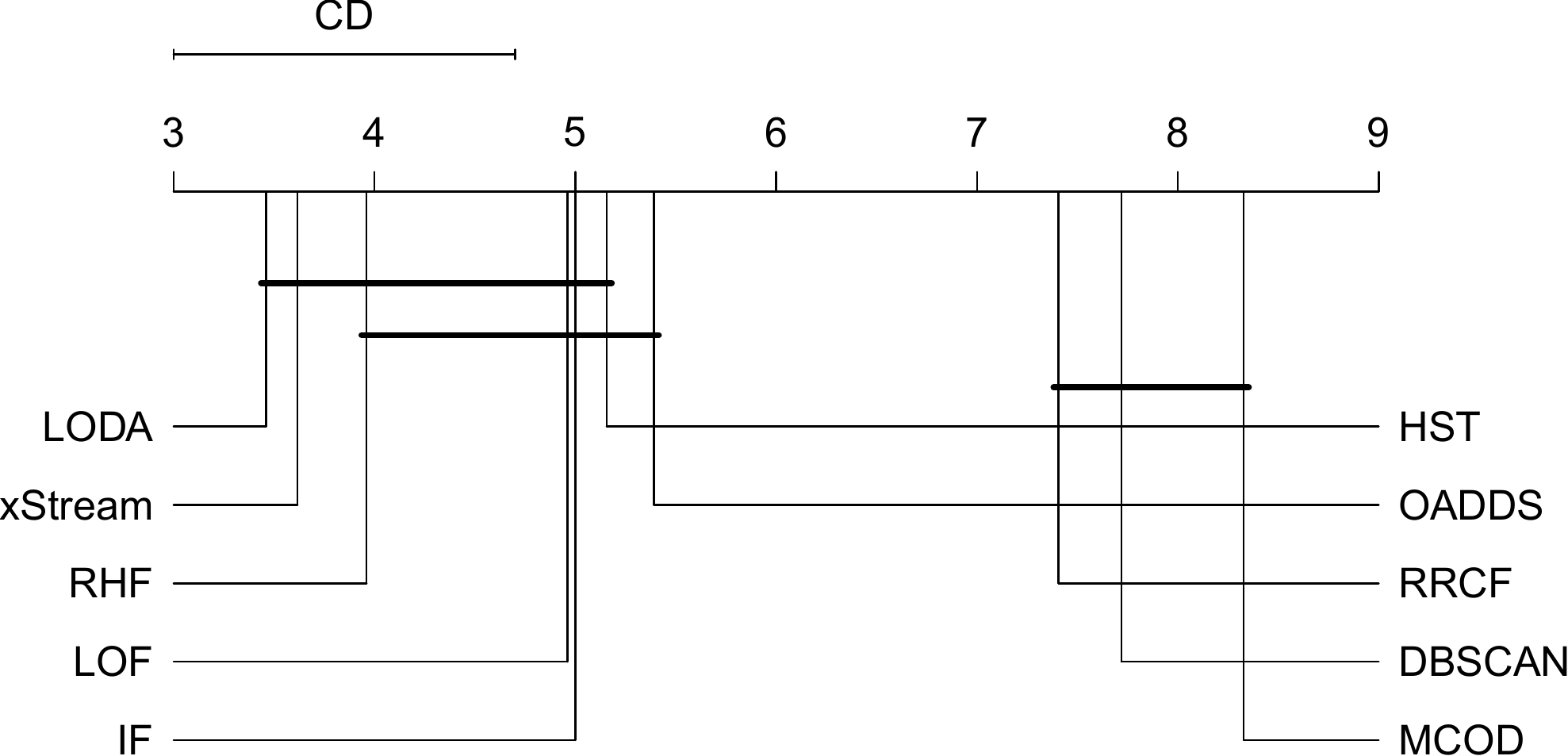}
   (a)$\qquad\qquad\qquad\qquad\qquad\qquad\qquad\qquad\qquad\qquad\qquad\qquad$(b)
    \caption{\textit{Selection of AD algorithm}: average rank comparison between algorithms. Algorithms linked by a horizontal line are not significantly different from each other.
    (a) Ideal upper bound and (b) Practical lower bound evaluation scenarios.}
    \label{fig:ad_2_best_ranks}
\end{figure*}

As in most cases,  algorithmic AD performance is highly dependent on hyperparameter settings.  As usually done in the literature, we study performance over multiple hyperparametrizations for each algorithm. We empirically set bounds of the grid search based on recommendation of each algorithm, that we summarize in Table \ref{tab:params}.   We remark that some of the algorithmic hyperparameters refer to absolute number of samples or features. However, given our datasets have extremely heterogeneous sizes, using an absolute value would introduce an uncontrolled variation in their effect, and as thus we opt to use proportional values:  every parameter that is expressed in terms of percentages  follows this convention. 
We additionally remark that we fixed a some non-numerical parameters (for LODA, OADDS, RRCF and xStream) that are reported as footnotes in the table.

Overall, the results on the set of 10 algorithms reported in the following  account for 1277 hyperparameter combinations, equivalent to 68 hours of computational time. We remark that the number of hyperparameter combinations tested is not uniform across algorithms (from as low as 1 to as much as 324), which is tied to  both the heterogeneous number of hyperparameters across algorithms, as well as the to the widely variable computational requirements (e.g., RRCF  is over two orders of magnitude slower than several of the other algorithms, making it difficult to explore a large grid for it). 
At the same time, we carefully verified that we introduce no bias in the grid selection, as we observe  no correlation between AD performance and either of (i) the size of the grid explored ($\rho = -0.35$), or (ii) the overall running time of the algorithm ($\rho = -0.36$).
For instance, we will see that DBSCAN performance are consistently bad over the  260 combinations explored, whereas RHF and LODA rank among the best performing algorithms despite only 4 combinations were tested.

\subsubsection{Ideal ensemble}

We start by a thorough study of AD performance, to gather in a scientifically principled way, a ideal upper-bound of algorithmic performance in academic settings. The performance of the \emph{Ideal ensemble} are outlined in Fig.\ref{fig:ad_1_best} for all datasets, with datasets ranked in order of increasing Pr-Rec AUC.    Recall that the ideal ensemble  (i) optimally tune  the  hyperparameter settings (i.e., those maximizing the Pr-Rec AUC )  for each algorithm and then (ii) selects a posteriori the best performing algorithm  (i.e., the one with the highest  Pr-Rec AUC). The Pr-Rec AUC for the ideal ensemble is reported as a dashed line in Fig.\ref{fig:ad_1_best}, that envelopes the individual algorithm selected by the ensemble, represented with colored bars. 

Notice that in about 1/2 of the datasets, multiple algorithms yield to the same results, in which case a gray bar is reported. Notice also that, among the other algorithms, xStream,  LODA  and IF appear consistently as being selected by the ideal ensemble.  Further, notice that the Pr-Rec AUC is higher than 0.99 (0.75) for 1/2  (3/4) of the datasets, with just very few   (1/8) datasets with Pr-Rc AUC lower than 0.5.

While performance are remarkable, the gap with the \emph{AD Oracle}  (i.e., the use of the ground truth given by expert annotation) is still evident. 
By cursory observation, we gather that root cause of this phenomenon is intrinsically tied to the nature of human labeling: (i) experts label some of the samples they feel important (e.g., as they have a business impact) but can omit to label all consecutive samples 
    (e.g., since they are interested in the timely detection of the event) or samples that are inconsequential (e.g., no  business impact) although such samples are objectively statistical outliers: this increases false alarm rate, and contribute in the decreasing the precision of the algorithm (and as such the Pr-Rec AUC);
    (ii) complementary, expert may be particularly attentive to label moments in time where some features, that have a specially important semantic to them, appear to have even slight variations: this instead increases the false negative rate, and decreases the recall of the algorithm (and as such the Pr-Rec AUC).

Notice that the above observations are intrinsically tied to the fact that we are assessing performance on a timeslot-per-timeslot basis: as such, this does not directly imply that anomalous ``events'', that often spans multiple timeslots, are necessarily unnoticed; similarly, when false positive AD samples immediately follow true positive detection, this  reduces the Pr-Rec AUC though this apparent performance degradation has no negative consequences in practice. Notice further that the use of Expert Knowledge (EK) is designed to tackle the problem (ii), by seamlessly encoding probabilistic information readily available from expert solutions.

\subsection{Selection of AD algorithm}\label{sec:ad:select}

The previous subsection illustrated academic  performance for the best possible case, i.e., an ideal ensemble,  able to use all algorithms and that can optimally tune the hyperparameters of each algorithm.  
In this section, we instead aim at understanding 
(i) the performance of each algorithm in isolation, 
(ii) measuring the distance from the ideal ensemble,  and 
(iii) recommend ``good enough'' algorithm with a single hyperparametrization.

\subsubsection{Ideal upper-bound (UB)}
We now consider each algorithm in isolation, but still allow to ideally tune the hyperparameters of each algorithm on each dataset, gathering an ideal upper-bound of each algorithm performance.

This is reported in Fig.~\ref{fig:ad_2_best}-(a) in the form of a scatterplot of the median Pr-Rec AUC of each algorithm against the median computation time per sample (in milliseconds). For reference, we also report performance of the ideal ensemble, whose computation time assumes parallel execution of each individual algorithm, and as such is dominated by the slowest algorithm\footnote{Clearly, the comparison between ensemble vs individual algorithms is not fair in terms of the raw amount of computation, that would be captured by the sum of the times.}
The picture allow to clearly cluster algorithms in the different quadrants, and to additionally factor in the computational cost of the AD module, which is important for practical reasons.  Remarkably, xStream stands out for being close to the Ideal Ensemble UB performance in median, followed by IF (which is also computationally simpler), LODA (on par)  and RHF (computationally more costly). 
As previously outlined, LODA and RHF performance are good even though the number of explored hyperparametrs is comparatively smaller.

To further analyze to what extent the observed differences are statistically significant, we measure the mean rank of the algorithm and perform a Nemenyi test\cite{rankcomparison}, a post hoc test that performs pair-wise comparisons, testing each algorithm against each other. Two classifiers' performances are considered significantly different if their average ranks differ by at least the Critical Difference, calculated as:
\begin{equation}
    CD=q_{\alpha}\sqrt{\frac{k(k+1)}{6N}}
\end{equation}
with $k$ being the number of algorithms to compare, $N$ the available number of samples and $q_{\alpha}$ a value based on the Studentized range statistic divided by $\sqrt{2}$ ($q_{\alpha} = 3.164$ for 10 algorithms and a significance of 0.05). We report these in Fig.~\ref{fig:ad_2_best_ranks}-(a). For each algorithm, the picture reports the 
average rank across all datasets; algorithms are sorted from lowest average rank (left and top) to highest rank (right and bottom). The resulting CD$=3.164\sqrt{\frac{10\cdot 11}{6 \cdot 64}} = 1.69$ is annotated in the picture. The figure further links algorithms whose rank distance is below the CD threshold, and that are thus not significantly different in a statistical sense. This analysis confirms that xStream, IF and LODA, if opportunely tuned, are all able to provide very good AD performance, with RHF and LOF close followers.

\subsubsection{Practical lower-bound (LB)}
We now put in place a further practical restriction, and require a single hyperparameter setting (listed in bold in Tab.\ref{tab:params}) for each algorithm across all datasets.
The Ideal Ensemble LB is subject to the same restrictions, as it is now able to select only among 10 instances (compared to 1277 previously for the UB case).

Interesting takeaways can be gathered from Fig.~\ref{fig:ad_2_best}-(b), which are further corroborated by the rank analysis reported in  Fig.~\ref{fig:ad_2_best_ranks}-(b). First, the performance degradation of the  Ideal Ensemble LB and of xStream remains limited. Second,  performance are now significantly more separated, with performance of proximity-based algorithms (DBSCAN, LOF, MCOD, OADDS), and of distance-based algorithms (DBSCAN, LOF) in particular, showing the largest degradation. Third, with the exception of RHF, performance of tree-based ensembles also degrades significantly (IF performance exhibit a sharp degradation), whereas performance of subspace-based ensembles (LODA, xStream) remains consistent.

These observations have important practical implications: first,  xStream, LODA ad RHF  are able to provide robust and good AD performance even in practical settings.
Second, while an ensemble is still desirable and recommended, a single algorithm can be expected to provide satisfactory performance.  In what follows we 
consider xStream, LODA ad RHF (the best performing in the extended algorithmic set)  and additionally include IF (the best performing in the limited set of the previous  conference version~\cite{itc32}).

\section{Feature Scoring and Expert Knowledge}\label{sec:results:FS}

\begin{figure}[t]
\begin{center}
  \includegraphics[width=0.75\columnwidth]{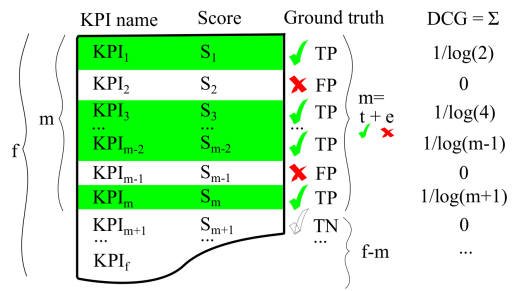}
  \caption{Synoptic of the performance metrics for attention focus mechanism.}
  \label{fig:fs:synoptic}
\end{center}
\end{figure}

Our aim in this section is to assess the performance of the 
FS module (contrasting na\"ive methods, to state of the art methods and our non-parametric proposals), as well as quantifying the benefits brought by simple statistical expert knowledge EK. For the sake of simplicity, we  restrain our attention to  a subset of  the most relevant anomaly detection algorithms identified in the previous section.
As in the previous section, we first define the performance metrics (Sec.\ref{sec:res:metrics_fs}), 
next overview FS and EK performance (Sec.\ref{sec:res:fsglance}), 
then dig deep into FS  (Sec.\ref{sec:res:fs}) and finally conservatively quantify EK benefits (Sec.\ref{sec:res:ek}).

 \begin{figure*}[t]
 \centering
  \includegraphics[width=0.4\textwidth]{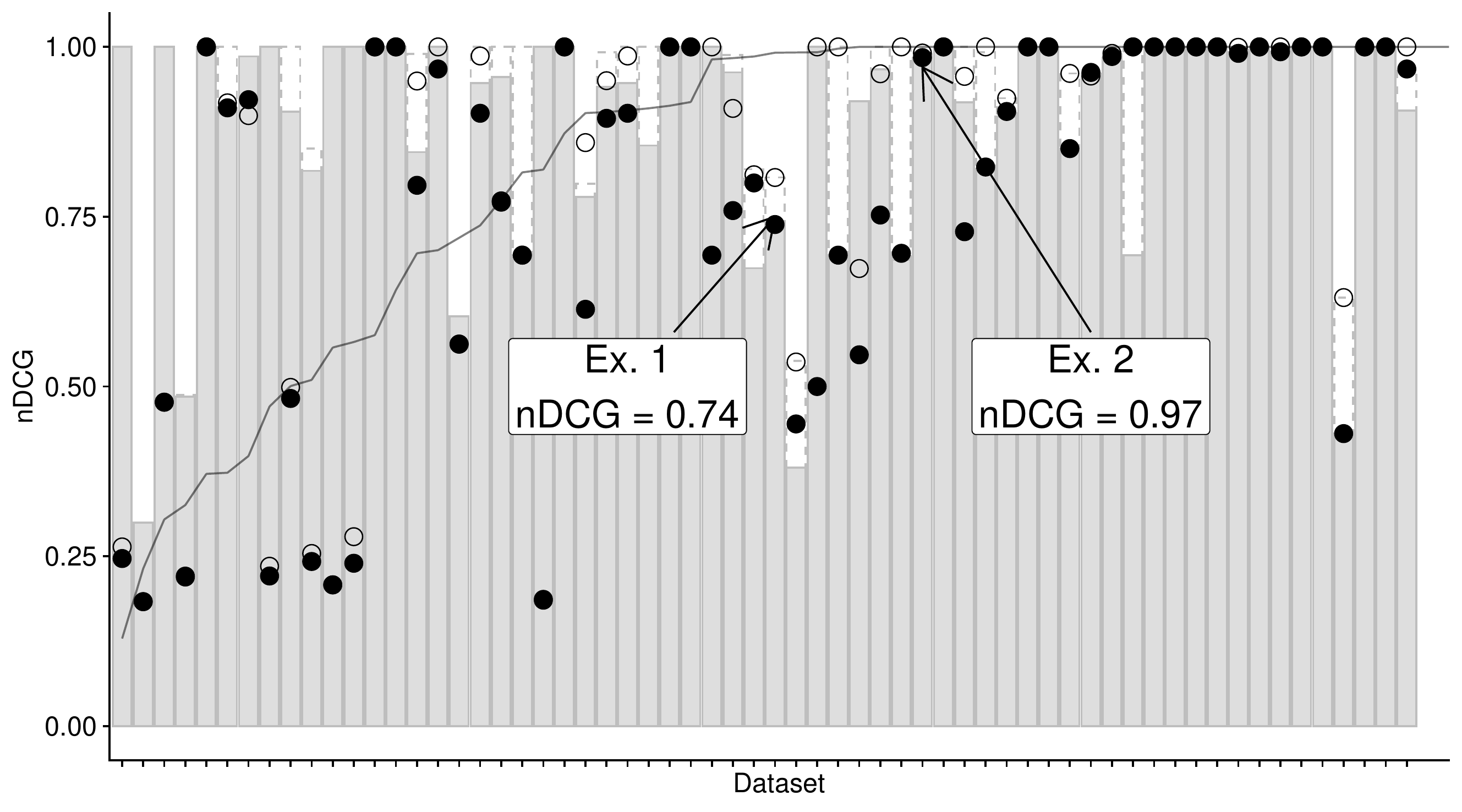}
  \raisebox{0.8cm}{\fbox{\includegraphics[width=0.28\textwidth]{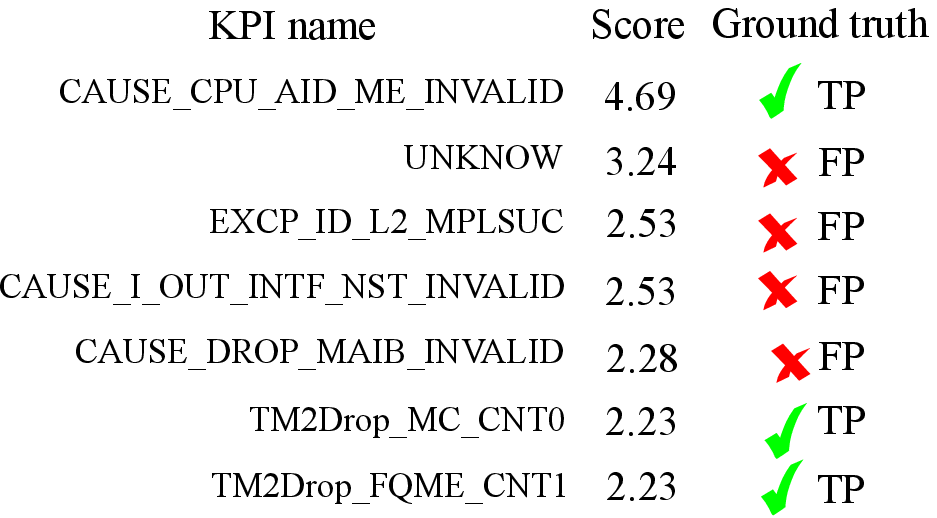}}}
  \raisebox{0.8cm}{\fbox{\includegraphics[width=0.28\textwidth]{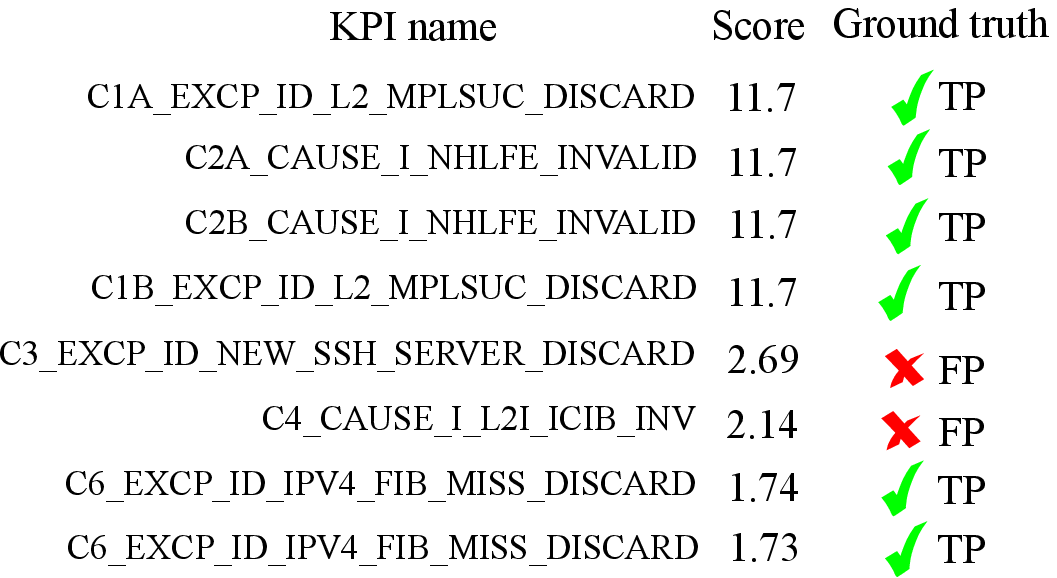}}}
  \caption{\textit{Experimental FS results at a glance}: nDCG  for oracle (grey bar) vs ensemble (black points) across all datasets (left) and  example sorting for two nDCG values (right). Dataset are ranked for increasing Pr-Rec AUC  as 
  in Fig. \ref{fig:ad_1_best} (corresponding Pr-Rec AUC shown in gray contour line): while nDCG  correlates with AUC, (i) it is still possible to obtain high FS nDCG despite poor AD Pr-Rec AUC and (ii) vice-versa obtain poor FS nDCG despite good AD.}
  \label{fig:fs:glance}
\end{figure*}

\subsection{Performance metrics}\label{sec:res:metrics_fs}
We define metrics to assess the extent of agreement among the ranking induced by $\hat s$ and the ground truth $g$, defined by the expert, that has observed and judged $f$ features to find the $t<f$ anomalous ones, with the goal of minimizing the number $e$ of irrelevant features that are presented to the expert.

\subsubsection{Network expert viewpoint}

Fig.~\ref{fig:fs:synoptic}  illustrates an example output where the $f$ features are sorted according to their decreasing anomalous score $s$ so that $s_i>s_j$ for $i<j$.
Ideally, an expert would want an algorithm to  present him these $t$ most relevant features at the top of the ranking. 
In particular, we  denote with $m = t + e$ the last position occupied in our ranking by a feature that the expert has flagged as relevant to troubleshoot the issue. By definition, all features occupying a rank higher than $m$ are non-anomalous (true negatives), whereas $e$ KPIs within the $m$ topmost returned by the ranking are false positives (feature 2 and $m-1$ in the example of Fig.~\ref{fig:fs:synoptic}). Formally we define the \emph{Reading effort}, $m = t+e$ as the number of KPIs $m$ that must be examined by an expert using HURRA to view all $t\le m$ anomalous KPI.

\subsubsection{Machine learning viewpoint}
From a machine learning viewpoint, a robust metric to measure the extent of agreement between the expert ground truth $g$ and the output ranking induced by $s$ is represented by the normalized Discounted Cumulative Gain (nDCG)  which is commonly used in information retrieval to compare rankings~\cite{yahoo_rankings_from_xgboost} and is defined as the ratio of two factors:
\begin{equation}
nDCG=DCG/iDCG
\end{equation}
The ideal DCG score (iDCG) is attained when all KPIs flagged as anomalous by the expert appear in the first $t$ positions (the order of non-anomalous KPIs after $t$ is irrelevant):
\begin{equation}
iDCG = \sum_{i=1}^{t<n}\frac{1}{log_2{(i+1)}}
\end{equation}
\noindent The DCG score of any ranking can be computed by accounting only for the anomalous KPIs:
\begin{equation}
DCG = \sum_{i=1}^{n}\frac{\mathbb{1}( \sum_t g_{it} > 0 )} {log_2{(i+1)}}
\end{equation}
\noindent where it should be noted that the KPI returned in the $i$-th position by a FS policy may not be present in the ground truth  (false positive in Fig.\ref{fig:fs:synoptic}) in which case it has null contribution to DCG, as well as it lowers contribution of the next true positive feature. nDCG is  bounded in [0,1] and equals 1 for perfect matching with the expert solution: although less intuitive, this metric allow to unbiased comparison across datasets, as well as quickly grasping the optimality gap of the HURRA solution.

\subsection{Performance at a glance}\label{sec:res:fsglance}
We start by illustrating  performance at a glance over all datasets in Fig.\ref{fig:fs:glance}.  The picture reports the nDCG scores obtained  using the best FS function across all datasets (ranked by increasing nDCG), when the AD task is performed by an Oracle (grey bar, upper bound of the performance) or by an  Ensemble of unsupervised methods (black points, realistic results)
with (light points and no shading) or without (dark points and shading) Expert knowledge.
It can be seen that, in numerous troubleshooting cases, the ensemble is able to attain high levels of agreement with the expert (the dots approach the envelope). It also appears that leveraging expert knowledge can assist both the ensemble as well as the Oracle, which is interesting. Finally, it can also be seen that there are few cases that are particularly difficult to solve (nDCG$<$0.25 in 3 cases, where the number of anomalous/total features is 1/157, 2/335 and 1/346 respectively): anomaly detection appears difficult for the algorithms we considered due to the curse of dimensionality, so subspace-based methods~\cite{mazel2011sub,sota_hics,sota_lookout} would be more appropriate in these cases (cfr. Sec.\ref{sec:discussion}).

To assist the reader with the interpretation of the nDCG metric, Fig.\ref{fig:fs:glance} additionally reports the ranking produced by HURRA in two examples datasets, that portray cases of moderate and high nDCG scores. 
For instance, in the high-nDCG case (Ex. 2), it can be seen that the $t=6$ anomalous out of $f=39$ total features are within a reading effort of $m=8$ KPIs with $e=2$ false positive for an overall nDCG$=0.97$. Similar observations can be made for the low-nDCG case (Ex. 1), where there is a higher number of false positive ($e=11$).

\subsection{Selection of Feature Scoring (FS) policy}\label{sec:res:fs}
To select a FS strategy, we proceed in two steps: first, we decouple FS performance from AD performance by leveraging the AD Oracle, contrasting several methods.
Second, we assess how the performance of the selected FS strategies are impacted by the use of real AD algorithms.

\subsubsection{Decoupled AD-FS performance} 

\begin{figure}[t]
 \centering
  \includegraphics[width=0.5\textwidth]{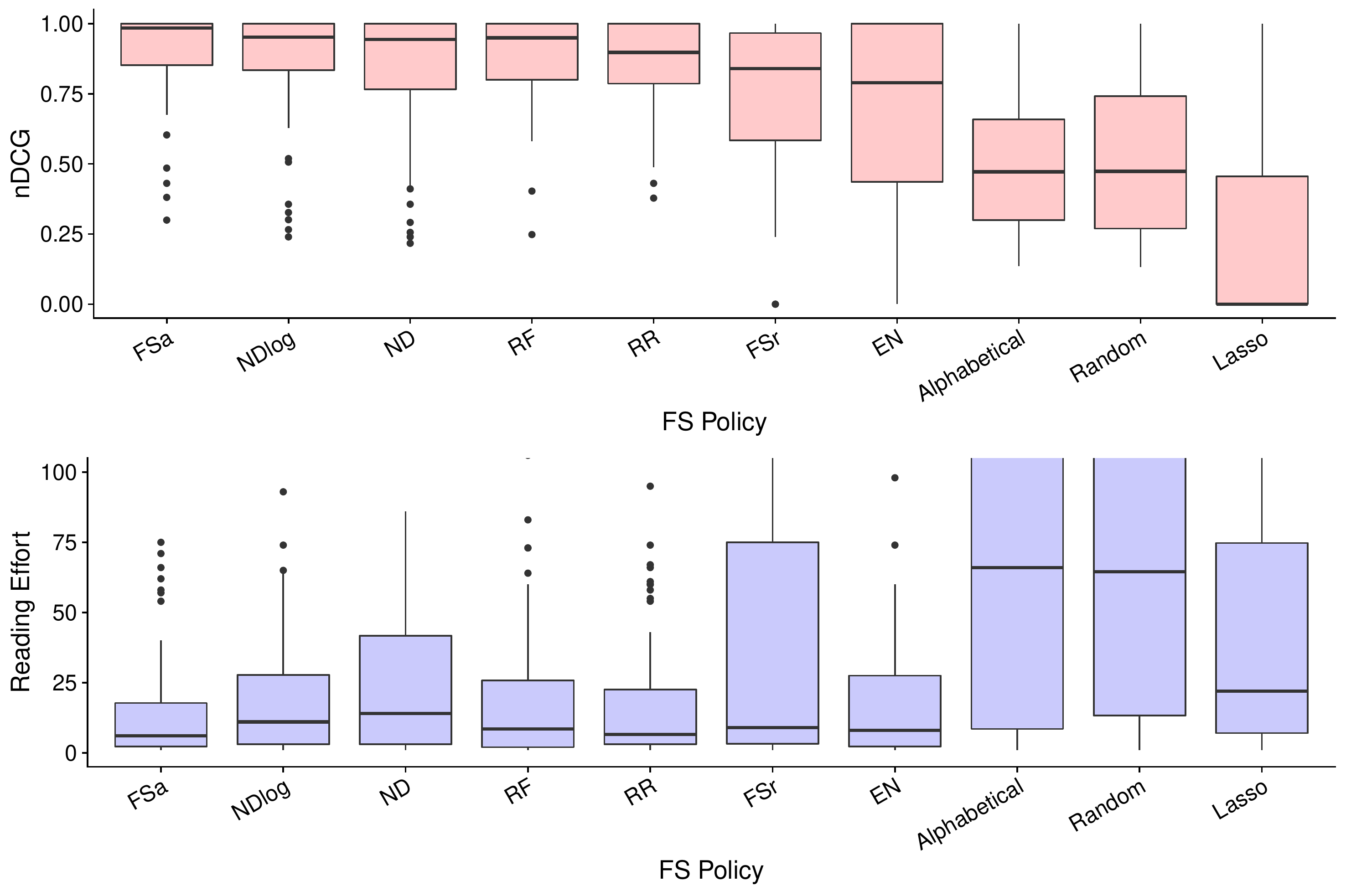}
  \caption{\emph{Decoupled AD-FS performance}: Impact of Feature Scoring policy on nDCG and reading effort metrics for ideal AD Oracle, sorted by decreasing median nDCG.}
  \label{fig:FS}
\end{figure}

\begin{figure}[t]
  \centering
  \includegraphics[width=0.5\textwidth]{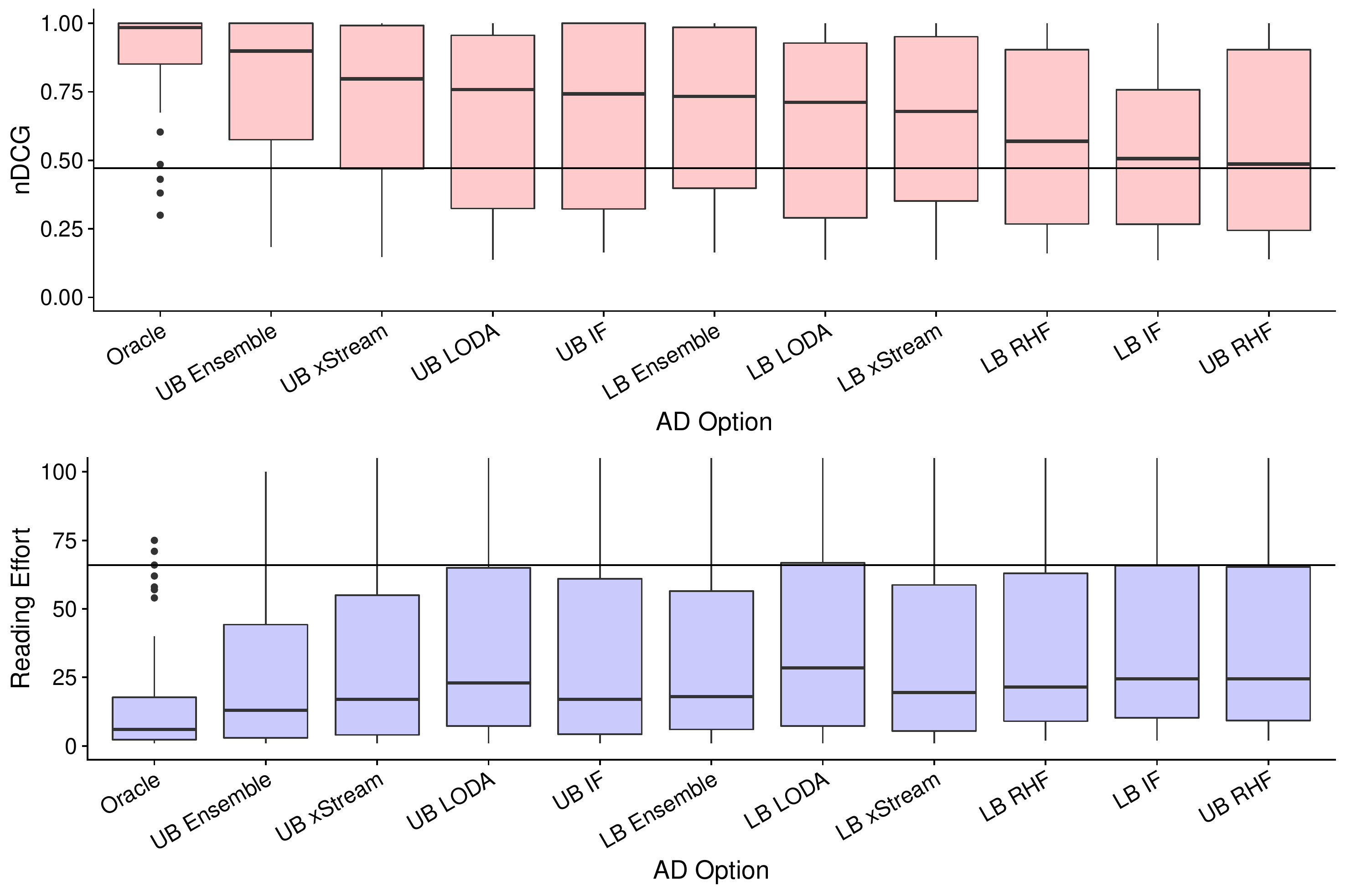}
  \caption{\emph{Coupled AD-FS performance}: Impact of Anomaly Detection on nDCG and reading effort on $FSa$ policy, sorted by decreasing nDCG. Horizontal reference line represents the median performance of an ideal AD Oracle with na\"ive Alphabetic FS ranking.}
  \label{fig:AD}
\end{figure}

Fig.\ref{fig:FS} next contrasts FS performance for nDCG and reading effort metrics considering 10 policies.
Two rankings are determined by the proposed non-parametrics average-based $FSa$  and  rank-based $FSr$ feature scoring defined by (\ref{eq:FSa}) and (\ref{eq:FSr}) respectively. Na\"ive rankings that are  evaluated as the theoretical expectation of \emph{Alphabetical} and  \emph{Random}. State of the art in parametric \emph{descriptive rankings}\cite{sota_other_adele} are represented by normal \emph{ND} and lognormal \emph{NDlog} strategies, whereas 
state of the art in \emph{predictive rankings}\cite{PROTEUS} are
represented by random forest \emph{RF}, or 
the   \emph{RR},   \emph{EN}   and   \emph{Lasso} variations of ElasticNet logistic regression.

It can be seen that (i) the proposed $FSa$ strategy outperforms all others, with  
(ii) descriptive $NDlog$ and  predictive $RR$ closely following; (iii)  additionally, $FSa$ ranking   increase nDCG by roughly a factor of two with respect to  alphabetical (or random) sorting: worth stressing from a network domain viewpoint is the fact that alphabetical and random sorting have close performance, which can be also interpreted as an indication that presenting KPIs in alphabetical order has no value for the troubleshooting expert. Also worth noting from a machine learning viewpoint (iv) $Lasso$ feature selection negatively impact nDCG: this can be explained with the fact that $Lasso$ tends to remove features that have high mutual information, whereas the expert may tend to label all such features, which explains the nDCG drop.

\begin{figure}[t]
\centering
  \includegraphics[width=0.5\textwidth]{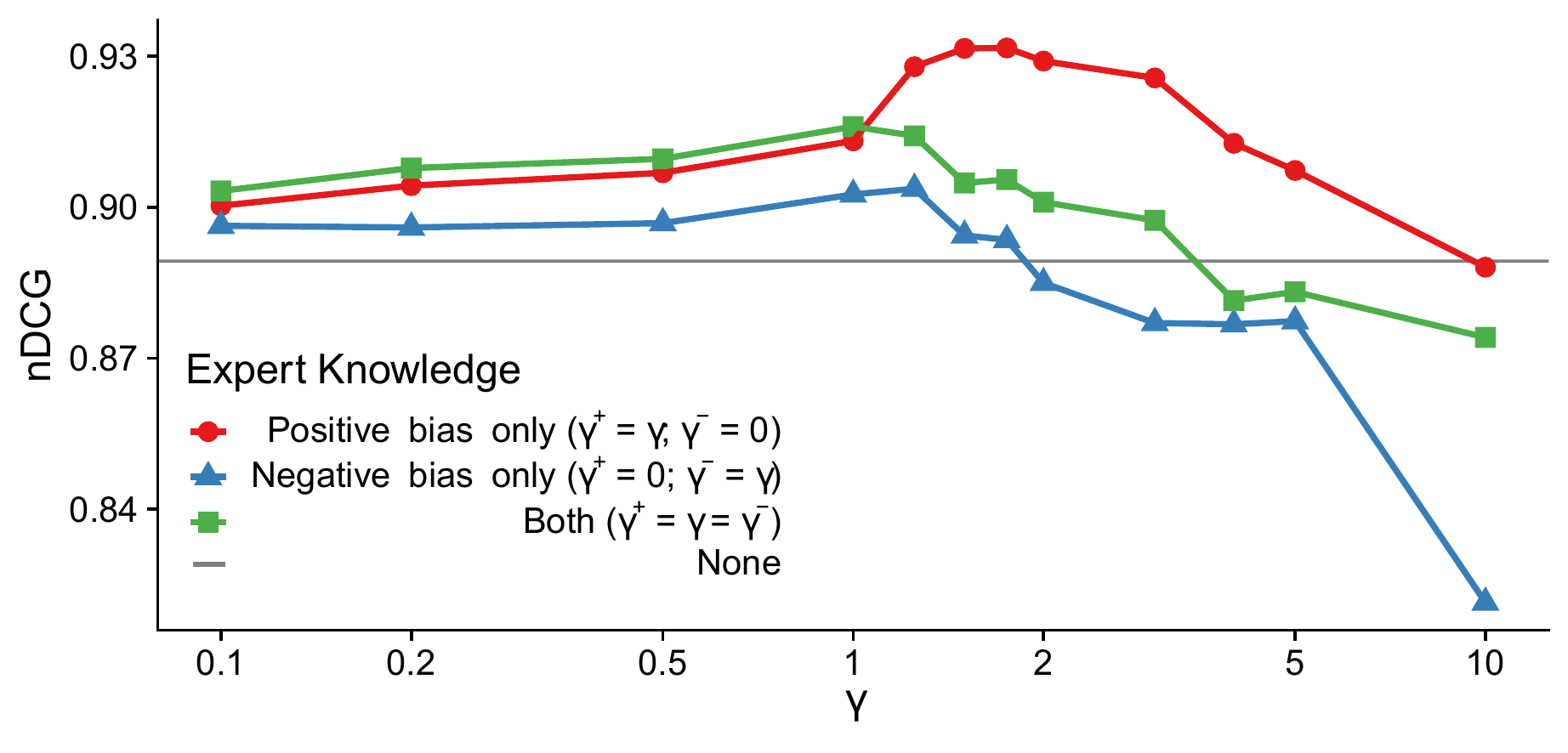}    \caption{\emph{Impact of Expert Knowledge}:  average nDCG over all datasets on the anomaly detection Oracle with FSa scoring. Comparison of negative, positive and combined FSa+EK (points) vs FSa-only (solid line).}    \label{fig:EK:oracle}
\end{figure}

\subsubsection{Coupled AD-FS performance}\label{sec:res:ad}
Limitedly considering the best performing average-based feature scoring $FSa$, we now consider  the impact of AD
choices in Fig.\ref{fig:AD}. Particularly, boxplots report the quartile of nDCG and reading effort for
Oracle, Ideal ensemble and the four AD algorithms previously selected (in UB and LB settings), for a total of 11 cases.   

Notice in particular that (i) the Oracle, Ensemble and xStream with $FSa$ provide performance that are quite close in median, as well as (ii) vastly better than an AD Oracle  with Alphabetical FS sorting  of the current legacy systems (horizontal reference line).
Notice also that in median, (iii) the reading effort for a very conservative practical setting (xStream LB with $FSa$)  remains below 20 features, compared to about 66 of the ideal AD Oracle  with Alphabetical FS sorting,
Overall, we see that simple practical feature scoring (FSa) brings a significant 3-fold reduction of work with respect to ideal anomaly detection, and even when using a single algorithm in practical settings (xStream LB).

\subsection{Impact of Expert Knowledge (EK)}\label{sec:res:ek}
We next gauge the value of the expert knowledge on the simple yet effective average-based $FSa$ feature scoring, using both AD  Oracle as well as real AD algorithms. 

\subsubsection{AD Oracle} 
Considering the Oracle first, the solid horizontal black line in Fig.\ref{fig:EK:oracle} corresponds to the average nDCG over all datasets gathered without the use of EK: clearly, improving such a high nDCG should prove quite difficult, and this should thus be considered as a conservative scenario to assess benefits deriving from EK.  
We consider several cases to better isolate the effects of positive and negative bias, particularly (i) positive only $\gamma^+=\gamma$ and $\gamma^- = 0$, (ii) negative only 
 $\gamma^+=0$ and $\gamma^- =\gamma$ and (iii) both effects with equal gain $\gamma^+= \gamma^- =\gamma$.
Several interesting remarks are in order. First, observe that improvements in the ranking are already evident for moderate gain $\gamma \ge 1/10$. Second, observe that positive bias $K^+_j$ has a stronger effect than negative one $K^-_j$, yet the effects bring a further slight benefit when combined. Third, the benefit of positive bias tops around $\gamma=2$ (in these datasets) and remains beneficial even for very large gains $\gamma\approx 10$.
Fourth, the effects of negative bias can instead worsen the resulting rank for $\gamma>2$ so that the combination of positive and negative bias (with equal weight) is also affected for large gains.

These different effects are easy to interpret considering the different nature of positive and negative biases. In particular, positive bias $K^+_j$  assesses the rate at which observers individually flag KPI $j$, as observed by independent observers over multiple independent datasets. Whereas $K^+_j$ does not directly translate into the probability that KPI $j$ is also anomalous in the dataset under observation, increasing the score proportionally to it  can help shifting the attention  toward KPIs that are often considered valuable toward the  troubleshooting solution.
Interestingly, as the the support of the observation grows, the metric becomes more accurate, and as the bias grows, more frequent problems can be solved more effectively. 
In contrast,  the computation of negative bias $K^-_j$ couples observations across several metrics, as the numerator in the  expression relates to the number of observations where the KPI $j$ is flagged as normal despite its score $s_j$ exceeds the score of at least one KPI flagged as anomalous. However, the subset of KPIs in common among any pair of dataset is small, so the knowledge distilled in $K^-_j$ appears more difficult to transfer across ISPs.

\begin{figure}[t]
\centering
  \includegraphics[width=0.5\textwidth]{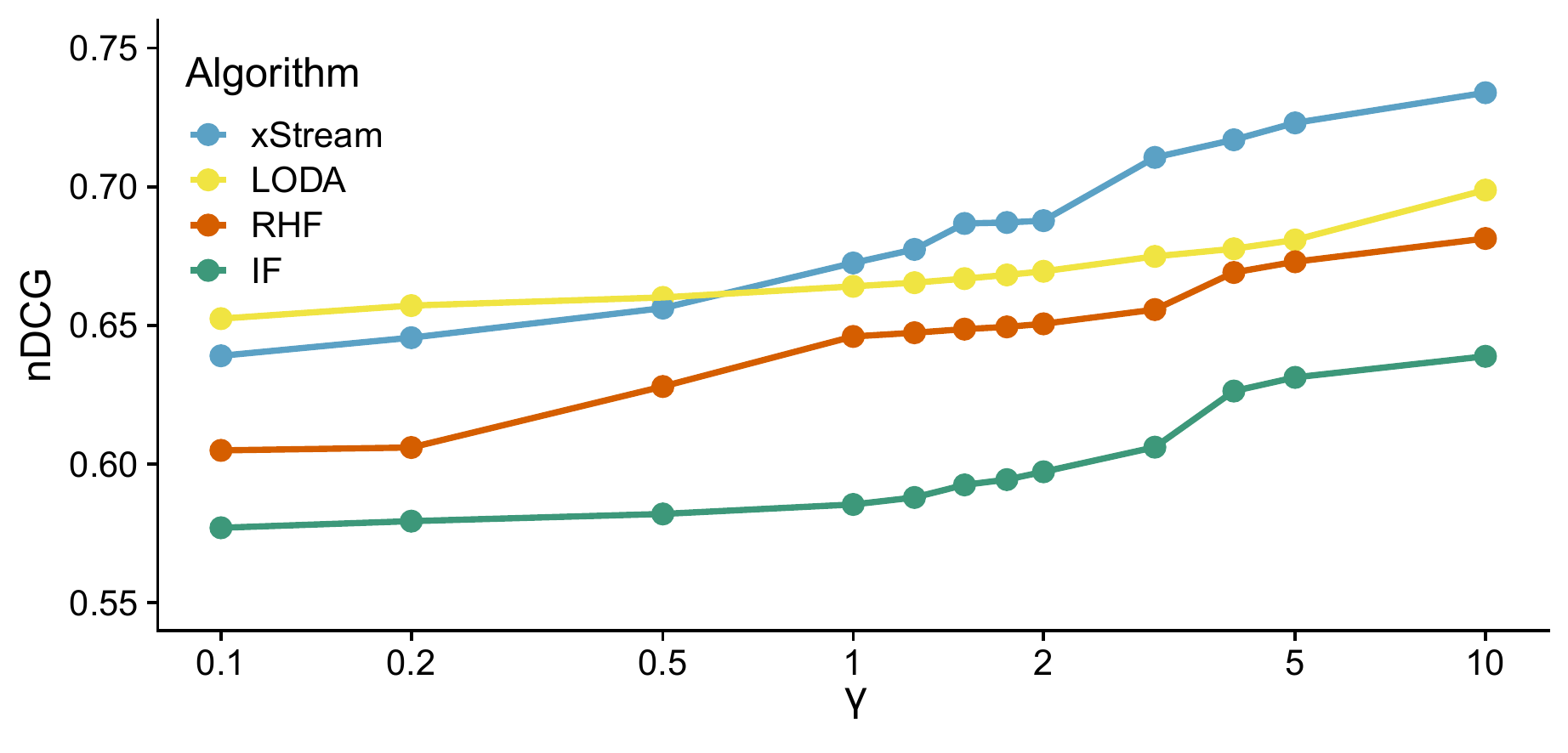}
    \caption{\emph{Impact of Expert Knowledge}: average nDCG over all datasets for the top-performing anomaly detection algorithms, using FSa scoring.}
    \label{fig:EK:algorithms}
\end{figure}

In summary, (i)  it appears that simple frequentist representation of EK are providing measurable advantages, even on top of oracle anomaly detection with high nDCG;
(ii) experimental results do not recommend the use of negative bias $K^-_j$, so that setting  $\gamma^-=0$ is adviseable;
(iii) when combined with an ideal AD Oracle, the gains from positive bias $K^+_j$ appears to saturate  around 
$\gamma\approx 2$.

\subsubsection{AD Algorithms} 

Finally,  we turn our attention to real AD algorithms in practical settings. Discarding the negative bias for the reasons just outlined, we limitedly focus on the the impact of positive bias EK on real AD algorithms in LB settings, which in shown in Fig.\ref{fig:EK:algorithms} averaged across all datasets.

It is interesting to notice that, in practical settings:  (i) the EK gain is qualitative consistent across algorithms,   (ii) the use of EK compensate for the lack of accuracy of AD algorithms with respect to the oracle
and (iii)  the gain from positive bias is  consistent for a very large parameter range $\gamma^+\in(0,10]$: selecting $\gamma^+>1$ can thus be expected to significantly help experts troubleshooting faster the most common problems.

\section{Full-blown System  Results}\label{sec:results:E2E}
Finally, we report the full system performance (Sec.\ref{sec:e2e:perf}) and discuss the limits of HURRA, by outlining open research points that are of interest for the whole community (Sec.\ref{sec:discussion}). 

\subsection{Performance at a glance}\label{sec:e2e:perf}
\begin{figure}[t]

  \includegraphics[width=0.48\textwidth]{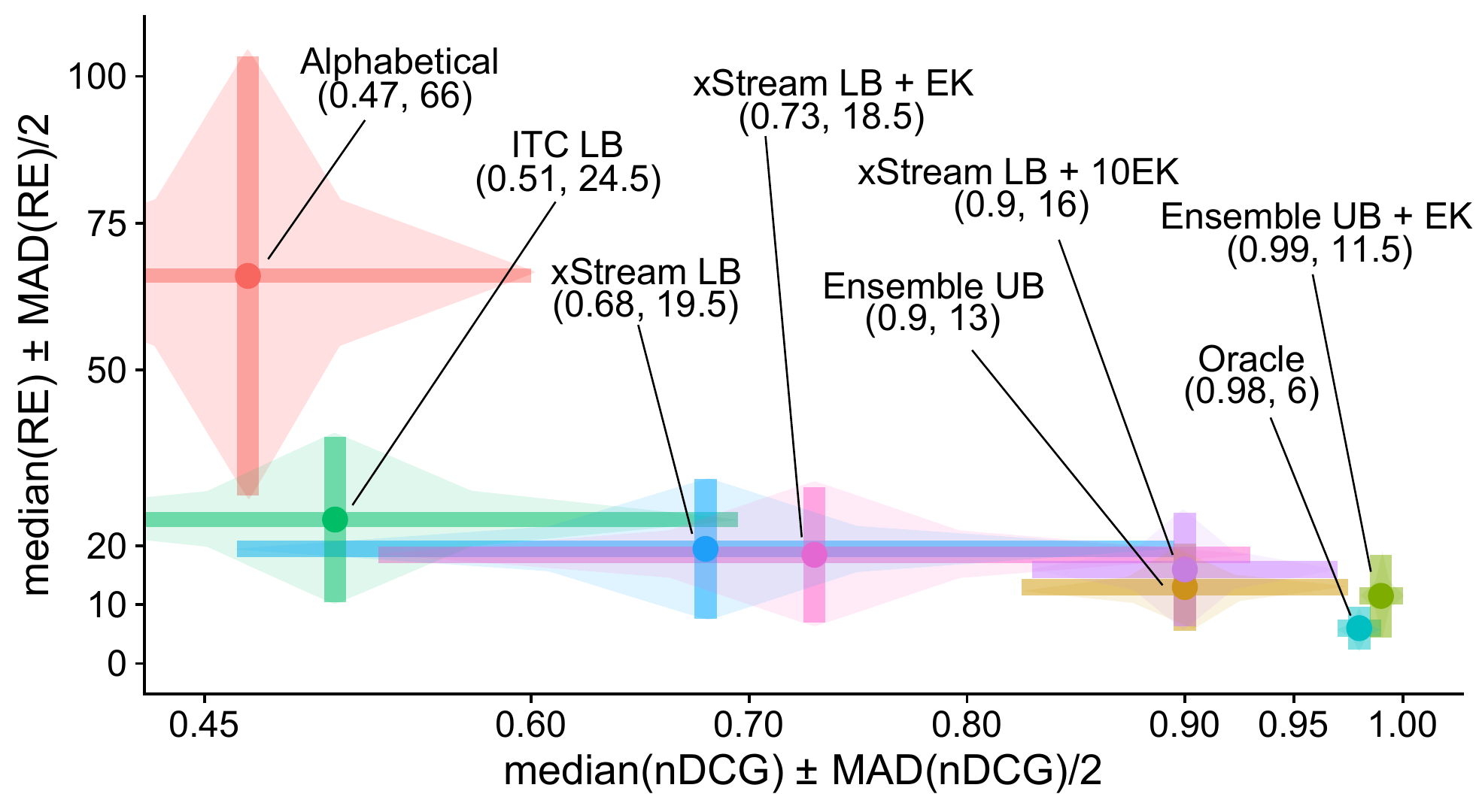}
  
 \vspace{8pt}
\begin{footnotesize}
\begin{tabular}{rllllp{1cm}}
  \toprule
   & {\bf  AD} & {\bf FS} & {\bf EK} $\gamma$ & {\bf nDCG} & {\bf RE} \\ 
  \midrule
     & Oracle & Alphabetical & 0 & 0.47 & 66 \\ 
\midrule
\parbox[t]{2mm}{\multirow{4 }{*}{\rotatebox[origin=c]{90}{\textbf{Practical}}}}
    & IF LB (ITC\cite{itc32}) & FSa & 0 & 0.51 & 24 \\ 
    & IF LB (ITC\cite{itc32}) & FSa & 10 & 0.73 & 19 \\ 
    & xStream LB  & FSa & 0 & 0.68 & 19 \\ 
    & xStream LB  & FSa & 10 & 0.90 & 16 \\ 
\midrule
\parbox[t]{2mm}{\multirow{4}{*}{\rotatebox[origin=c]{90}{\textbf{Ideal}}}}
 & Ensemble UB & FSa & 0 & 0.90 & 13 \\ 
 &  Oracle & FSa & 0 & 0.98 & 6 \\ 
 & Ensemble UB  & FSa & 10 & 0.99 & 8 \\ 
 & Oracle  &  FSa & 2 & 1.00 & 4 \\ 
\bottomrule
\end{tabular}

  \caption{\textit{Full blown system results}. Relative impact of HURRA's building blocks as a scatter plot of median nDCG vs Reading effort (RE) performance (ideal performance in the bottom right corner). Performance are further tabulated in detail, showing that the use of expert knowledge can (xStream LB $\gamma=10$) can attain ideal ensemble performance (Ensemble UB). Further, expert knowledge is also beneficial in the ideal Ensemble/Oracle cases.}
  \label{fig:relative}

\end{footnotesize}

\end{figure}

We compactly summarize performance of the whole HURRA system as a scatter plot in Fig.\ref{fig:relative}.
Performance of the AD Oracle with  na\"ive  Alphabetical system appears on the top-left corner (reference baseline, corresponding to a reading effort of 66 KPIs), far from the bottom-right corner of AD Oracle with FSa scoring (ideal target, corresponding to a reading effort of 6 KPIs, a 10-fold improvement).  A practical HURRA system using average feature scoring $FSa$  is able to achieve interesting operational points even with  single algorithm and single hyperparametrization:  IT LB (best in  ITC\cite{itc32}) and xStream LB reduce reading effort to 24 and 19 respectively.  The use of Expert Knowledge (EK) further allow a practical HURRA system (xStream LB with $\gamma=10$) to approach the ideal performance of Ensemble UB algorithm. 

Finally, EK appears to be beneficial even for ideal algorithms, although with diminishing returns. The EK gain even in the presence of an oracle for anomaly detection can be explained with the fact that not all KPIs with a manifest anomalous behavior are important, as they may be a symptom rather than a cause. This  confirms that even a simple frequency-based approach in exploiting expert knowledge can lead the expert in focusing on KPIs that are closer to the ``root cause'', without directly exploiting causality, and that this information can be ``transferable'' in machine learning terms, across datasets.
Overall, these results confirms the interest for exploiting, in a semi-supervised way, even simple statistical expert knowledge, since EK efficiently compensates for weak performance  of unsupervised anomaly detection algorithms.

\subsection{Discussion}\label{sec:discussion}
While performance benefits of FS and EK HURRA's building block are clear, even on top of ideal AD algorithms, the limits of HURRA are also worth pointing out, as they open interesting research for the community.

\subsubsection{Auto Anomaly Detection (AutoAD)} Our results show that, while it is possible to deploy relatively weak AD algorithms even with a single hyperparametrization, results could be improved by combining results of multiple algorithms, or by using different hyperparametrizations.
In particular, while the use of statistical Expert Knowledge (EK) as proposed in this paper  is an effective way to cope with the lack of accuracy of individual AD algorithms/hyperparametrizations, it does not represent however a  general solution, as e.g,, it requires supervision and would be hardly transferable across domains.
Devising a completely unsupervised AutoAD strategy, able to approach performance of the ideal ensemble  by comparing results of randomized trials is a promising direction requiring future research.

\subsubsection{Stream-mode feature scoring (StreamFS)}  While this work considers stream-mode anomaly detection algorithms (unlike the previous conference version~\cite{itc32}, that only considered batch anomaly detection methods), the feature scoring evaluation is still performed in a batch fashion.
Since human operators are interested in immediately reacting to faults, an open research point is to further elucidate the tradeoff between \emph{timeliness} (e.g., 
evaluating FS solely on the base of the first detected anomalous sample) and \emph{accuracy} (i.e., degradation of nDCG metrics), to push the boundaries of stream-mode interpretability in anomaly detection.

\section{Conclusions}\label{sec:conclusions}

This paper presents HURRA, a system designed to reduce the time taken by human operators for network troubleshooting. HURRA is simple and modular, with a design that decouples the Anomaly Detection (AD),   Feature Scoring (FS) and Expert Knowledge (EK) blocks.
HURRA not only assists human operators in their tasks (using unsupervised learning techniques for AD and FS), but that can also learn from the final troubleshooting solutions validated by the operator (without requiring his explicit interaction 
to seamlessly build  an EK knowledge base).  HURRA is easily deployable (as it requires minimal changes with respect to UI and no human interaction), hot-swappable (as it can benefit but does not require an EK base) and is expected to improve over time (as the EK base grows).

 Our  performance evaluation shows that whereas single algorithms AD with fixed hyperparametrization do provide weak anomaly detection perfomance, the use of an additional FS block has the potential of making unsupervised systems of practical values, by greatly reducing the troubleshooting effort.
 Additionally, our results  show  that the use of EK block can completely
 compensate the lack of accuracy of AD algorithms, by attaining in practice the same performance of ideal AD ensembles (e.g.,  using multiple algorithms and  hyperparametrization), which makes HURRA highly appealing from a practical viewpoint.


\bibliographystyle{IEEEtran}
\bibliography{refs} 

\begin{IEEEbiography}[{\includegraphics[width=1in,height=1.25in,clip,keepaspectratio]{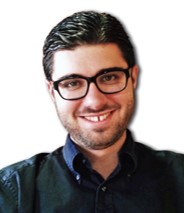}}]{José Manuel Navarro}
 is a Senior Research Engineer on Network AI at Huawei Technologies Co.,Ltd. He received his MSc from Universidad Miguel Hernández de Elche in 2013 and his PhD in 2018 from Universidad Politécnica de Madrid and was a visiting researcher at National Institute of Informatics in Tokyo, Japan, during 2016. He has coauthored 2 patents and 10+ papers on Anomaly Detection, Network Failure Prediction and Explainability. He is a Member of IEEE.
\end{IEEEbiography}

\begin{IEEEbiography}[{\includegraphics[width=1in,height=1.25in,clip,keepaspectratio]{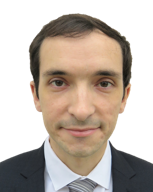}}]{Alexis Huet}
 is a Senior Research Engineer on Network AI at Huawei Technologies Co., Ltd. He received his MSc from ENS Lyon in 2010 and his PhD in 2014 from Claude Bernard Lyon 1 University. He then worked at Nanjing Howso Technology (2015-2018) as a data scientist, where he was responsible for data mining, modeling and implementation for different machine learning projects. His main research interests include computational statistics and applied mathematics for machine learning. He is a member of IEEE.
\end{IEEEbiography}

\begin{IEEEbiography}[{\includegraphics[width=1in,height=1.25in,clip,keepaspectratio]{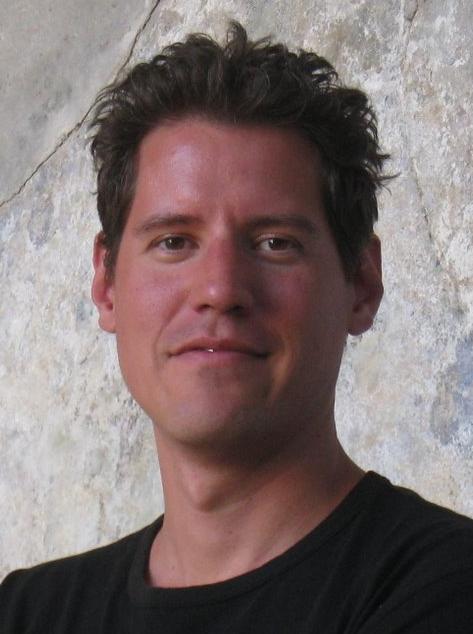}}]{Dario Rossi}
is  Network AI CTO and Director of the DataCom Lab  at Huawei Technologies, France. Before joining Huawei in 2018, he was Full Professor at Telecom Paris and Ecole Polytechnique and holder of Cisco's Chair NewNet@Paris.  He has coauthored 15+ patents and 200+ papers in leading conferences and journals, that received 10 best paper awards, a Google Faculty Research Award (2015) and an IRTF Applied Network Research Prize (2016).  He is a Senior Member of IEEE and ACM.
\end{IEEEbiography}

\end{document}